\documentclass{applemlr}

\usepackage{amsmath}
\usepackage{enumerate}
\usepackage{algorithm}
\usepackage{algpseudocode}
\usepackage{amsfonts}
\usepackage{amsthm}
\usepackage{cleveref}
\usepackage{diagbox}
\usepackage{colortbl}
\usepackage{amssymb}
\usepackage{xspace}
\usepackage{wrapfig}
\usepackage{adjustbox}
\usepackage{tabularx}
\usepackage{booktabs}
\usepackage{mathtools}
\usepackage{tikz}
\usepackage{enumitem}
\usepackage{silence}
\usepackage{dsfont}
\usepackage[table]{xcolor}
\usepackage[dvipsnames]{xcolor}
\usepackage{multirow}
\usepackage{makecell}
\usepackage{xfakebold}

\usepackage{amsmath,amsfonts,bm}

\def\eqref#1{equation~\ref{#1}}

\def\1{\bm{1}}

\DeclareMathAlphabet{\mathsfit}{\encodingdefault}{\sfdefault}{m}{sl}
\SetMathAlphabet{\mathsfit}{bold}{\encodingdefault}{\sfdefault}{bx}{n}

\definecolor{textgray}{HTML}{6E6E73}
\usetikzlibrary{positioning, calc}
\usetikzlibrary{decorations.pathmorphing}

\makeatletter
\patchcmd{\wrong@fontshape}{\@gobbletwo}{}{}{}
\makeatother
\numberwithin{equation}{section}
\makeatletter
\AtBeginDocument{
  \urlstyle{sf}
  
}
\makeatother

\definecolor{light}{RGB}{125, 125, 125}
\crefname{tcb@cnt@pbox}{code}{code}
\Crefname{tcb@cnt@pbox}{Code}{Code}
\crefname{assumption}{assumption}{assumption}
\Crefname{assumption}{Assumption}{Assumptions}

\newtcolorbox[auto counter]{pbox}[2][]{
  colback=white,
  title=Code~\thetcbcounter: #2,
  #1,fonttitle=\sffamily,
  fontupper=\sffamily,
  arc=2pt,
  colframe=bgcolor,
  coltitle=fgcolor,
  colbacktitle=bgcolor,
  toptitle=0.25cm,
  bottomtitle=0.125cm
}

\makeatletter
\newcommand\applefootnote[1]{%
  \begingroup
  \renewcommand\thefootnote{}%
  \renewcommand\@makefntext[1]{\noindent##1}%
  \footnote{#1}%
  \addtocounter{footnote}{-1}%
  \endgroup
}
\makeatother

\definecolor{cverbbg}{gray}{0.90}

\usepackage[utf8]{inputenc}
\usepackage{graphicx}
\usepackage{caption}
\usepackage{array}

\title{Adversarial Distilled Retrieval-Augmented Guarding Model\\for Online Malicious Intent Detection}

\author{
Yihao Guo\textsuperscript{1}, 
Haocheng Bian\textsuperscript{1}, 
Liutong Zhou\textsuperscript{2,†}, 
Ze Wang\textsuperscript{1}, 
Zhaoyi Zhang\textsuperscript{3,†}, \\
Francois Kawala\textsuperscript{1}, 
Milan Dean\textsuperscript{5,†}, 
Ian Fischer\textsuperscript{1}, 
Yuantao Peng\textsuperscript{1}, 
Noyan Tokgozoglu\textsuperscript{1}, \\
Ivan Barrientos\textsuperscript{1}, 
Riyaaz Shaik\textsuperscript{1}, 
Rachel Li\textsuperscript{1}, 
Chandru Venkataraman\textsuperscript{1}, 
Reza Shifteh Far\textsuperscript{1}, \\
Moses Pawar\textsuperscript{1}, 
Venkat Sundaranatha\textsuperscript{1}, 
Michael Xu\textsuperscript{1}, 
Frank Chu\textsuperscript{4,†} \\

}

\affiliation{
\textsuperscript{1}Apple \quad
\textsuperscript{2}Cohere \quad
\textsuperscript{3}DeepMind \quad
\textsuperscript{4}Meta \quad
\textsuperscript{5}MongoDB \quad
\textsuperscript{†}Work done while at Apple
}

\abstract{
With the deployment of Large Language Models (LLMs) in interactive applications, online malicious intent detection has become increasingly critical. However, existing approaches fall short of handling diverse and complex user queries in real time. To address these challenges, we introduce ADRAG (Adversarial Distilled Retrieval-Augmented Guard), a two-stage framework for robust and efficient online malicious intent detection. In the training stage, a high-capacity teacher model is trained on adversarially perturbed, retrieval-augmented inputs to learn robust decision boundaries over diverse and complex user queries. In the inference stage, a distillation scheduler transfers the teacher’s knowledge into a compact student model, with a continually updated knowledge base collected online. At deployment, the compact student model leverages top-$K$ similar safety exemplars retrieved from the online-updated knowledge base to enable both online and real-time malicious query detection. Evaluations across ten safety benchmarks demonstrate that ADRAG, with a 149M-parameter model, achieves 98.5\% of WildGuard-7B's performance, surpasses GPT-4 by 3.3\% and Llama-Guard-3-8B by 9.5\% on out-of-distribution detection, while simultaneously delivering up to 5.6$\times$ lower latency at 300 queries per second (QPS) in real-time applications.
}

\metadata[Correspondence]{\sffamily  mjockers@apple.com}
\date{\sffamily 18 September 2025}

\begin{document}

\maketitle

\section{Introduction}

Online malicious intent detection is increasingly critical as more Large Language Models (LLMs) are deployed in real-time interactive systems, where users may misuse them to produce non-compliant or harmful content \citep{OpenAI2023,Liu2023AgentBench,Han2024WildGuard}. For instance, users have successfully jailbroken an LLM-powered music application to generate racist songs \citep{MusicAbuse2024}. Detecting malicious intent in user inputs is therefore essential for ensuring safe and responsible LLM-based applications.

Traditional approaches for malicious intent detection fall into three categories: 
\textit{Supervised classifiers}, including deep networks and LLMs (e.g., BERT) trained on annotated safety datasets~\citep{Wulczyn2017}, 
are straightforward but suffer from rigid taxonomies, noisy labels, and weak generalization; 
\textit{General-purpose LLMs}, which perform classification through safety-oriented prompts~\citep{Bai2022,Zhang2024SCoT}, 
are more flexible but remain vulnerable to adversarial inputs; 
and \textit{Dedicated safety LLMs}, trained on safety-specific data and enhanced with techniques such as deliberative alignment~\citep{Guan2024DeliberativeAlignment}, 
achieve stronger alignment but are typically designed for non-RAG settings and often overlook retrieval-induced risks.

Beyond these limitations, most traditional methods are designed for offline use and cannot effectively support \textbf{real-time, context-aware intent detection} due to architectural and latency constraints. For example, a music application query such as \textit{``songs like `I Deserve to Bleed'"}, intended for search and recommendation, may be incorrectly flagged as unsafe if the model lacks knowledge that \textit{``I Deserve to Bleed''} is a song title. In practice, enabling such capabilities requires overcoming two main challenges: (1) lightweight models may meet latency requirements but often lack the reasoning and generalization needed for nuanced or adversarial queries; and (2) effective moderation increasingly requires dynamic, input-specific context that static classifiers cannot handle.

In this paper, we present \textbf{ADRAG} (Adversarial Distilled Retrieval-Augmented Guard), a unified framework for accurate, robust, and efficient online malicious intent detection. In the training stage, a high-capacity teacher model is trained on adversarially perturbed, retrieval-augmented inputs to learn context-aware classification under adversarial conditions. In the inference stage, a distillation encoder scheduler transfers the teacher’s knowledge to a compact student guard model, whose capabilities are continually refined through an online knowledge base. At deployment, the student model retrieves top-$K$ safety examples from the knowledge base to enable real-time, context-aware malicious intent detection.

This paper makes three main contributions:  
\begin{enumerate}[leftmargin=0.8cm, labelsep=0.3em]
    \item  We introduce ADRAG, the first framework that unifies adversarial training, context-augmented modeling, and selective knowledge distillation for real-time malicious intent detection. 
\item  We demonstrate that ADRAG, with a 149M-parameter model, matches the performance of GPT-4, WildGuard-7B, and Llama-Guard-3-8B across ten safety benchmarks, while sustaining an online latency under 6\,ms at 300 QPS.
    \item We present ablation studies demonstrating that Retrieval-Augmented Adversarial Fine-Tuning (RAFT) and Selective Knowledge Distillation (SKD) constitute complementary pillars of ADRAG, jointly enabling state-of-the-art performance in accuracy, robustness, and efficiency.

\end{enumerate}

\section{Related Work} 

\paragraph{Traditional Malicious Intent Detection}  
Existing approaches to malicious intent detection can be broadly categorized into supervised classifiers, general-purpose LLMs, and dedicated safety LLMs. \textbf{Supervised classifiers} include systems such as Perspective API~\citep{PerspectiveAPI2017}, OpenAI’s Moderation API~\citep{OpenAIModerationAPI2023}, rule-based approaches like RBE~\citep{Clarke2023RBE}, and task-adaptive models such as STAND-Guard~\citep{Wang2024STANDGuard}. \textbf{General-purpose LLMs} employ safety-oriented prompting strategies, such as Constitutional AI~\citep{Bai2022} and reflective prompting with SCoT~\citep{Zhang2024SCoT}. \textbf{Dedicated safety LLMs} are refined through techniques like behavior unlearning~\citep{Zhou2023Unlearnable}, and deliberative alignment~\citep{Guan2024DeliberativeAlignment}, with representative models including Llama-Guard~\citep{Inan2023}, ShieldGemma~\citep{Zeng2024ShieldGemma}, WildGuard~\citep{Han2024WildGuard}, BingoGuard~\citep{Yin2025BingoGuard}, and PolyGuard~\citep{Kumar2025PolyGuard}. Despite these advances, existing approaches are primarily designed for non-RAG settings and overlook the risks introduced by external retrieval.

\paragraph{RAG for Malicious Intent Detection}
Retrieval-augmented generation (RAG) enhances grounding and robustness by incorporating external context~\citep{Lewis2020RAG}. Beyond generation, recent work has extended RAG for safety tasks. Class-RAG~\citep{Chen2024ClassRAG} fine-tune Llama-3-8b with retrieved examples to adapt to changing moderation policies, while GuardAgent~\citep{Xiang2024GuardAgent} leverages retrieval to enforce safety constraints in planning agents. However, retrieved content may introduce harmful or misleading information~\citep{An2025RAGNotSafer}. While RAG improves adaptability, its retrieval component introduces new latency challenges. To address this,\textbf{ }Knowledge distillation~\citep{Hinton2015Distill} trains compact \textit{student} models to approximate the behavior of larger \textit{teacher} models. This approach has enabled efficient NLP models such as DistilBERT~\citep{Sanh2019DistilBERT}, and MiniLM~\citep{Wang2020MiniLM}. In the AI safety domain, distillation has been applied to build lightweight systems such as HarmAug~\citep{LeeEtAl24_HarmAug}, which distills Llama-3-8B into a 435M-parameter model. Our work combines RAG’s adaptability with distillation’s efficiency to deliver accurate, robust, and controllable malicious intent detection for real-time deployment.

\section{Preliminaries}

\subsection{Large Language Models}
 
Large Language Models (LLMs) $F_{\theta}(\cdot)$ are high-capacity neural networks that model the distribution of natural language. 
Given an input token sequence $\mathbf{x} = (x_1, \ldots, x_n)$ (e.g., a query or document), the model predicts the conditional probability of the output token sequence $\mathbf{y} = (x_{n+1}, \ldots, x_T)$ via an autoregressive objective:
{
 \begin{equation}
F_{\theta}(\mathbf{y} \mid \mathbf{x}) = \prod_{t=n+1}^{T} p(x_t \mid x_{<t}, \mathbf{x}),
 \end{equation}
}
where $\theta$ denotes the model parameters, $T$ is the total sequence length, and $x_{<t} = (x_{n+1}, \ldots, x_{t-1})$ represents all generated tokens before position $t$. The term $p(x_t \mid x_{<t}, \mathbf{x})$ denotes the conditional probability of token $x_t$ given its left context and the original input.

\subsection{Retrieval-Augmented Generation}
 
To mitigate the limitations of static knowledge and hallucinations in LLMs, Retrieval-Augmented Generation (RAG) integrates external knowledge by retrieving relevant information from a knowledge base $\mathcal{D} = \{\mathbf{d}_1, \ldots, \mathbf{d}_N\}$ at inference time, where $\mathbf{d}_{i}$ denotes an entity (e.g., a document or query-label pair). Given an input query $\mathbf{x}$, a target context set $\mathcal{D}_\mathbf{x} = \{ \mathbf{d}_{\mathbf{x}}^{(1)}, \ldots, \mathbf{d}_{\mathbf{x}}^{(K)} \}$ is constructed via a retrieval function:
{

\begin{equation}
\mathcal{D}_\mathbf{x} = \mathcal{R}_K(\mathbf{x}; \mathcal{D}),
\end{equation}
}
where $\mathcal{R}_K(\cdot)$ selects the top-$K$ relevant documents based on similarity. Here, $K$ is a hyperparameter controlling how many entries to retrieve. In practice, retrieval is often performed using approximate nearest neighbor (ANN) search over vector embeddings: $\underset{\mathbf{d}_i \in \mathcal{D}}{\text{TopK}} \;\text{sim}(f_{\text{enc}}(\mathbf{x}), f_{\text{enc}}(\mathbf{d}_i))$, where $f_{\text{enc}}(\cdot)$ is a shared encoder for queries and documents, and $\text{sim}(\cdot, \cdot)$ is a similarity metric such as cosine similarity. The LLM then generates output conditioned on both the input and the retrieved context: $F_{\theta}(\mathbf{y} \mid \mathbf{x}, \mathcal{D}_\mathbf{x})$.

\subsection{Adversarial Training}

Adversarial training enhances model robustness by optimizing performance under worst-case perturbations during training. Given an input sequence $\mathbf{x}$ and its corresponding target $\mathbf{y}$, a small perturbation $\delta$ is introduced into the input space, constrained by $\|\delta\| \leq \rho$, where $\rho > 0$ defines the perturbation budget. The training objective minimizes the worst-case negative log-likelihood under such perturbations:
{

\begin{equation}
\min_{\theta} \; \mathbb{E}_{\mathbf{x}, \mathbf{y}} \left[ \max_{\|\delta\| \leq \rho} -\log F_{\theta}^{\delta}(\mathbf{y} \mid \mathbf{x}) \right],
\end{equation}
}
where $F_{\theta}^{\delta}$ denotes the model’s output distribution conditioned on the perturbed input. This min-max formulation acts as a regularization mechanism, encouraging the model to learn smooth and locally stable representations around the training data manifold.

\subsection{Knowledge Distillation}
 
To enable efficient inference, a compact student model $F_{\theta_S}$ is trained to approximate the predictive behavior of a high-capacity teacher model $F_{\theta_T}$ under retrieval-augmented inputs. Given input $\mathbf{x}$ and retrieved context $\mathcal{D}_\mathbf{x}$, the distillation objective minimizes the Kullback–Leibler (KL) divergence between their output distributions:
{

\begin{equation}
\mathcal{L}_{\text{distill}} = \mathrm{KL}\left(F_{\theta_T}(\cdot \mid \mathbf{x}, \mathcal{D}_\mathbf{x}) \,\|\, F_{\theta_S}(\cdot \mid \mathbf{x}, \mathcal{D}_\mathbf{x})\right),
\end{equation}
}
where $\mathcal{L}_{\text{distill}}$ denotes the loss function measuring divergence between the teacher and student predictions, and $\mathrm{KL}(\cdot \,\|\, \cdot)$ is the Kullback–Leibler divergence. $F_{\theta_T}$ and $F_{\theta_S}$ represent the teacher and student models, parameterized by $\theta_T$ and $\theta_S$, respectively.  
In practice, two main approaches exist: Offline distillation freezes $\theta_T$ and trains the student on static teacher outputs, offering simplicity but often yielding suboptimal performance~\citep{Gou2021Survey,Kim2021SelfKD}. Online distillation updates both models jointly, enabling dynamic supervision at the cost of greater training complexity~\citep{Wang2020MiniLM}.

\subsection{Problem Statement}
 
\paragraph{Online Malicious Intent Detection}
Detecting user intent is critical for large-scale applications, where regulations such as the EU Digital Services Act~\citep{EUDSA} mandate proactive moderation of harmful or illegal queries. In these online platforms, real-time response is essential to ensure a seamless user experience. A natural approach is to deploy lightweight models fine-tuned on labeled intent datasets, which satisfy latency constraints but often underperform on rare, evolving, or context-sensitive queries. Maintaining high precision in these cases requires frequent retraining or manual updates, introducing scalability and maintenance challenges. 
To address these limitations, retrieval-augmented classification has emerged as a promising alternative~\citep{Buonocore2025RAR}. By retrieving policy-aligned examples from a curated knowledge base, these systems enhance generalization. Unlike static fine-tuned models, RAG-based architectures enable dynamic adaptation to novel queries without retraining, thereby enhancing flexibility and generalization.

Formally, RAG-based malicious intent detection is framed as a low-latency binary classification task predicting whether a user query $\mathbf{x}$ expresses malicious intent, conditioned on retrieved context $\mathcal{D}_\mathbf{x}$:
{

\begin{equation}
\hat{y} = F_{\theta}(\mathbf{x}, \mathcal{D}_\mathbf{x}), \quad 
\mathcal{T}_{\text{tot}}(\mathbf{x}) = \mathcal{T}_{\text{ret}}(\mathbf{x}) + \mathcal{T}_{\text{inf}}(\mathbf{x}, \mathcal{D}_\mathbf{x}) \leq \tau,
\end{equation}
}
where $\mathcal{T}_{\text{tot}}$ denotes the total system latency, comprising retrieval latency $\mathcal{T}_{\text{ret}}$ and inference latency $\mathcal{T}_{\text{inf}}$, which need to remain below a strict threshold $\tau$ to enable real-time moderation. To support context-aware prediction, the query $\mathbf{x}$ is concatenated with $K$ retrieved examples: $[\mathbf{x}; \mathbf{d}_{\mathbf{x}}^{(1)}; \ldots; \mathbf{d}_{\mathbf{x}}^{(K)}]$, where each $\mathbf{d}_{\mathbf{x}}^{(i)} = (\bar{\mathbf{x}}_i, \bar{y}_i)$ represents a prior query and its associated intent label. The concatenated input is fed into the classifier $F_\theta$ (or $\mathbf{x}$ alone if $\mathcal{D}_\mathbf{x}=\emptyset$), with prompt engineering used to ensure a coherent sequence that preserves semantic clarity and guides model behavior. This context-enhanced formulation improves reliability by leveraging semantically similar examples, while maintaining the strict latency constraints required for real-time applications.

\section{Adversarial Distilled Retrieval-Augmented Guard (ADRAG)}

\begin{figure*}[ht]
    \centering
    \includegraphics[width=\linewidth]{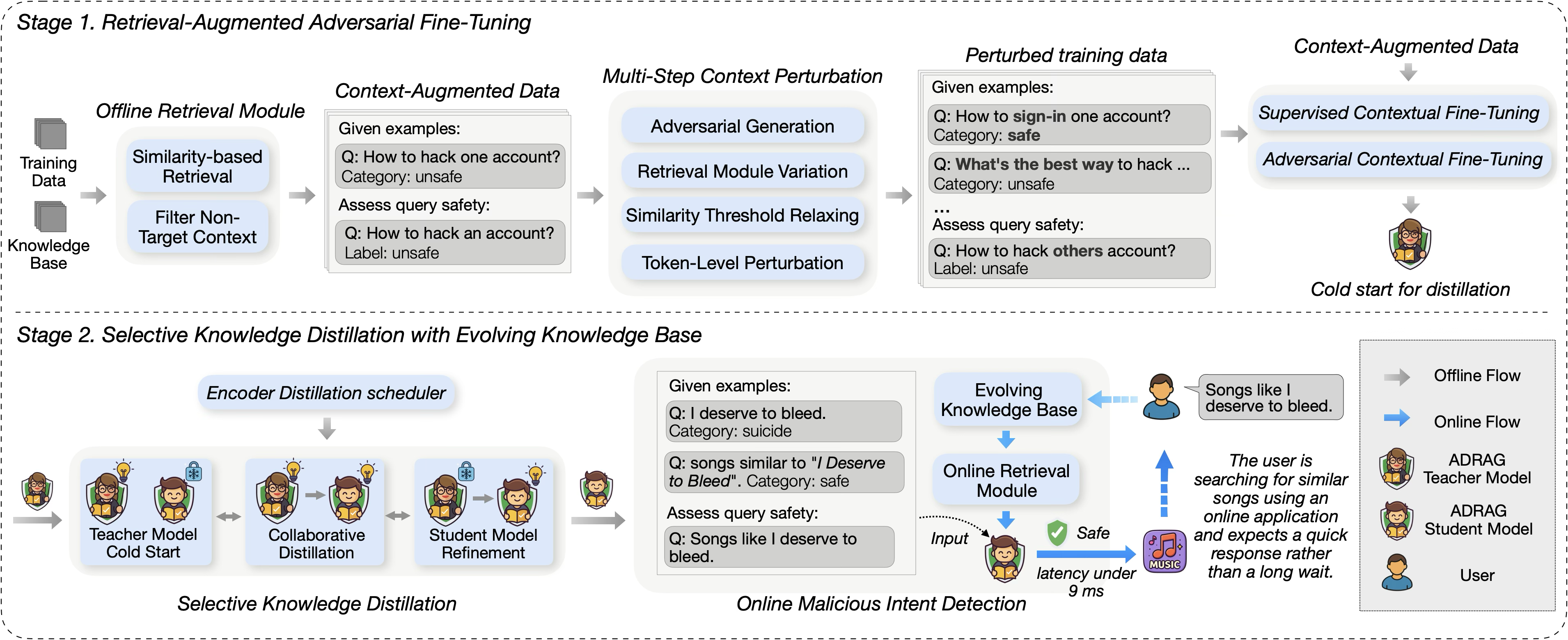}
    \caption{ADRAG Architecture}
    \label{fig:adrag_architecture}
\end{figure*}

\subsection{Overview}
 
We propose ADRAG, a two-stage framework for \textbf{Accurate}, \textbf{Robust}, and \textbf{Efficient} online malicious intent detection, as illustrated in Figure~\ref{fig:adrag_architecture}. In the training phase, a high-capacity teacher model $F_{\theta_T}(\mathbf{x}, \mathcal{D}_{\mathbf{x}})$ is optimized via \textbf{Retrieval-Augmented Adversarial Fine-Tuning (RAFT)}, where each input $\mathbf{x}$ is paired with a retrieved context set $\mathcal{D}_{\mathbf{x}}$ that injects external, often domain-specific knowledge to enrich under-specified queries, substantially enhancing the model’s predictive accuracy, especially for ambiguous or information-sparse cases. To further enhance robustness, ADRAG simulates real-world noise in the knowledge base, retrieval pipeline, context quality, and user input through adversarial perturbations, incorporating them into training via an adversarial loss $\mathcal{L}_{\text{adv}}$ that promotes consistent and reliable predictions under degraded conditions. In the inference phase, ADRAG applies \textbf{Selective Knowledge Distillation (SKD)} to compress the teacher into a lightweight student model $F_{\theta_S}$, guided by an encoder scheduler that governs training dynamics by assigning each epoch to operate in teacher-only, co-distillation, or student-focused mode. At deployment, the student retrieves top-$K$ relevant contexts $\mathcal{D}_{\mathbf{x}} \subset \mathcal{D}$ for each query $\mathbf{x}$ to enable accurate, low-latency predictions that generalize beyond the static training distribution and remain resilient to adversarial or ambiguous inputs, while adaptability is maintained through an\textbf{ Evolving Knowledge Base (EKB)} continuously updated with user feedback and policy-driven synthetic examples.

\subsection{Retrieval-Augmented Adversarial Fine-Tuning}
 
The training phase begins by equipping the teacher model with context-aware classification capabilities using retrieved external knowledge, followed by targeted adversarial training to enhance robustness against real-world data noise. While retrieval-augmented inputs improve prediction accuracy by enriching under-specified queries, they also introduce vulnerabilities: model outputs can become highly sensitive to perturbations in either the user query $\mathbf{x}$ or the retrieved context set $\mathcal{D}_{\mathbf{x}}$. A single misleading context $\mathbf{d}_{\mathbf{x}}^{(i)} \in \mathcal{D}_{\mathbf{x}}$, if contradictory to the ground-truth label $y_i$, can destabilize the model’s predictions. For example, a benign query such as ``I don't like the black color'' may mistakenly retrieve the toxic context such as ``I don't like black color people,'' whose racist connotation misleads the model's classification. These failures often stem from lexical overlap, retrieval mechanisms that prioritize surface similarity over intent alignment, or deeper semantic ambiguities \citep{An2025RAGNotSafer,Zou2024PoisonedRAG,Yoran2023RetRobust}. To mitigate these issues, RAFT trains the model to maintain consistent predictions even when the user query or retrieved context is adversarial or misleading.

\subsubsection{Supervised Contextual Fine-Tuning}
 
To enable the model to leverage informative, high-similarity context, supervised contextual fine-tuning is applied. Given a labeled dataset $\mathcal{D}_{\text{train}} = \{ (\mathbf{x}_i, y_i) \}_{i=1}^{N}$ and a knowledge base $\mathcal{D} = \{\mathbf{d}_1, \ldots, \mathbf{d}_N\}$, where each entry contains a query-label pair, a retrieved context set $\mathcal{D}_{\mathbf{x}_i} \subset \mathcal{D}$ is constructed for each input $(\mathbf{x}_i, y_i)$ using a similarity-based retrieval function: 
{

\begin{equation}
    \mathcal{D}_{\mathbf{x}_i} = \left\{ \mathbf{d} \in \mathcal{R}_K(\mathbf{x}_i; \mathcal{D}) \;\middle|\; \text{sim}(f_{\text{enc}}(\mathbf{x}_i), f_{\text{enc}}(\mathbf{d})) \geq 1 - \epsilon \right\},
\end{equation}
}
where $\epsilon$ controls the similarity tolerance for non-target contexts. The teacher model $F_{\theta_T}$ is then optimized using the following objective:
{
\begin{equation}
\mathcal{L}_{\text{train}} = \frac{1}{N} \sum_{i=1}^N \text{CE}\left(F_{\theta_T}(\mathbf{x}_i, \mathcal{D}_{\mathbf{x}_i}), y_i\right),
\end{equation}
}
where CE denotes the cross-entropy loss, defined as $\text{CE}(\hat{p}, y_i) = -\log \hat{p}_{y_i}$, which penalizes low predicted probability for the true class $y_i$.

\subsubsection{Multi-Step Context Perturbation.}

After acquiring context-aware classification capabilities, the model is further strengthened through exposure to perturbations that simulate diverse sources of noise commonly observed in real-world online systems. Specifically, each retrieved context set $\mathcal{D}_{\mathbf{x}_i}$ is augmented with adversarial variants to form an expanded set $\mathcal{D}_{\mathbf{x}_i}^\Delta$. These perturbations are designed to reflect variability across four critical steps of the intent detection pipeline: knowledge base construction, retrieval mechanisms, similarity-based filtering, and user-side input noise.

\paragraph{Adversarial Knowledge Base Generation}
In practice, a model's knowledge base may evolve over time or contain incomplete and noisy entries. To simulate such imperfections, a generative attack model $F_{\theta_A}$ is fine-tuned to produce adversarial variants for each retrieved context $\mathbf{d}_{\mathbf{x}_i} \in \mathcal{D}_{\mathbf{x}_i}$, as detailed in Appendix~\ref{sec:attack_model_ft}. These synthetic documents $\mathbf{d}_{\mathbf{x}_i}^\Delta \in \mathcal{D}_{\mathbf{x}_i}^\Delta$ may contradict the original label or introduce ambiguity, thereby challenging the model’s ability to generalize beyond clean, curated content.

\paragraph{Encoder and Similarity Variation}
Contemporary online systems often employ dense, sparse, or hybrid retrievers, resulting in substantial variability in retrieved context depending on the underlying configuration and domain-specific requirements. To reflect this variability, alternative encoders $f_{\text{enc}}^\Delta$ and similarity functions $\text{sim}^\Delta$ are introduced, including commonly used metrics such as cosine similarity, dot product, and lexical matching. Using these alternatives, a perturbed context set is constructed as follows:
{
\setlength{\abovedisplayskip}{0pt} 
\begin{equation}
\mathcal{D}_{\mathbf{x}_i}^\Delta = \mathcal{R}_K(\mathbf{x}_i; \mathcal{D}, f_{\text{enc}}^\Delta, \text{sim}^\Delta),
\end{equation}
}
which exposes the model to semantically diverse retrievals and encourages it to learn robust representations that generalize across heterogeneous retrieval configurations.

\paragraph{Similarity Threshold Relaxing}
Although retrieval systems typically apply a similarity threshold $\epsilon$ to filter out unrelated documents, real-world embeddings are often noisy and may assign high similarity scores to marginally relevant content. To simulate this behavior, the threshold is relaxed to include documents within a broader similarity range:
{

\begin{equation}
\mathcal{D}_{\mathbf{x}_i}^\Delta = \left\{ \mathbf{d} \in \mathcal{D} \;\middle|\; \delta \leq \text{sim}(f_{\text{enc}}(\mathbf{x}_i), f_{\text{enc}}(\mathbf{d})) \leq 1 - \epsilon \right\},
\end{equation}
}
where $\delta < \epsilon$ allows for lower similarity matches. This encourages the model to remain stable even when the retrieved context is only loosely related to the input query.

\paragraph{Sampling Perturbation}
To simulate user-side noise and informal language patterns, token-level perturbations such as character modifications, word reordering, and minor textual corruption are applied to both the input query $\mathbf{x}$ and its associated retrieved context set $\mathcal{D}_{\mathbf{x}_i}$. While these edits preserve the semantic intent, they introduce variation in surface form, encouraging the model to learn intent-aware representations that remain robust under noisy or imperfect input conditions.

\subsubsection{Adversarial Contextual Fine-Tuning}
 
Given the perturbed context sets $\mathcal{D}_{\mathbf{x}_i}^\Delta$, an adversarial loss is introduced to penalize inconsistent predictions under noisy or misleading conditions, formalized as:
\begin{equation}
\mathcal{L}_{\text{adv}} = \frac{1}{N} \sum_{i=1}^N \text{CE}\left(F_{\theta_T}(\mathbf{x}_i, \mathcal{D}_{\mathbf{x}_i}^\Delta), y_i\right),
\end{equation}

and the final training loss combines standard and adversarial components:
\begin{equation}\label{eq_total_loss}
\mathcal{L}_{\text{total}} = \mathcal{L}_{\text{train}} + \lambda \cdot \mathcal{L}_{\text{adv}},
\end{equation}

 where $\lambda$ is a weighting factor that balances accuracy and robustness. In practice, both the original and perturbed context sets, $\mathcal{D}_{\mathbf{x}_i}$ and $\mathcal{D}_{\mathbf{x}_i}^\Delta$, are pre-retrieved, generated, and merged with each query in the training set $\mathcal{D}_{\text{train}}$ before model optimization begins.

This Retrieval-Augmented Adversarial Fine-Tuning strategy enables the model to leverage accurate contextual signals, remain resilient to noisy or misleading inputs, and maintain consistent predictions under perturbations, thereby enhancing both accuracy and robustness in real-world deployment scenarios.

\subsection{Selective Knowledge Distillation with Evolving Knowledge Base}
 
Low-latency intent detection, a crucial requirement for real-time applications, motivates the use of compact models; however, as shown in Appendix~\ref{sec:small_model_raft}, compact models often struggle to capture context-dependent patterns and generalize to unseen domains. 
Knowledge distillation partially mitigates this gap by transferring knowledge from a well-trained larger teacher model to a compact student model; however, in context-augmented settings, this method often fails to preserve contextual reasoning, and student models may degrade without continuously updated knowledge. To address these challenges, ADRAG introduces a unified deployment framework combining \textbf{Selective Knowledge Distillation (SKD)} with an \textbf{Evolving Knowledge Base (EKB)}. SKD enables controllable and adaptive knowledge transfer through an encoder scheduler that gradually shifts supervision from teacher to student, enhancing transfer effectiveness. Meanwhile, EKB supplies real-time, query-specific context to support continual adaptation. Together, these components ensure accurate and efficient malicious intent detection in rapidly changing environments.
 
\subsubsection{Selective Knowledge Distillation}
 
Traditional knowledge distillation faces a fundamental tradeoff between stability and adaptability. Offline distillation relies on fixed teacher outputs, providing stable knowledge transfer but often failing to capture
subtle, context-dependent, or adversarial intent.  In contrast, online distillation introduces dynamic between teacher and student, improving adaptability but reducing controllability and destabilizing training, especially when the teacher is under-trained or over-fitted. To address this limitation, ADRAG introduces an \textbf{encoder scheduler}, a dynamic mechanism that governs the training process by assigning supervision modes to each epoch, defined as
{

\begin{equation}
\mathbf{S}_{\text{train}} = [s_1, s_2, \dots, s_T], \quad s_t \in \{0, 1, 2\},
\end{equation}
}
where $\mathbf{S}_{\text{train}}$ specifies models update mode at each epoch: mode 0 updates only the teacher, mode 1 updates only the student (with the teacher frozen), and mode 2 updates both jointly. The schedule follows a three-stage curriculum that ensures robust teacher initialization, adaptive knowledge transfer, and stable convergence. Transition points $T_1$ and $T_2$ divide training into teacher initialization ($1 \leq t < T_1$), collaborative distillation ($T_1 \leq t < T_2$), and student refinement ($T_2 \leq t \leq T$).

\paragraph{Teacher Initialization}
In the first stage with $s_t = 0$, the scheduler activates only the teacher, optimizing it with the loss in Eq.~\ref{eq_total_loss}. The student remains inactive to avoid premature exposure to unstable supervision signals, reducing noise propagation and establishing a stable foundation for knowledge transfer.

\paragraph{Collaborative Distillation}
Once the teacher reaches sufficient maturity, the scheduler transitions to $s_t = 2$, enabling joint updates of teacher and student. During this phase, the teacher continues task-specific and adversarial optimization, while the student aligns with the teacher’s evolving predictions. As both models are updated at each step, their parameters become step-dependent, denoted by $\theta_T^{(b)}$ and $\theta_S^{(b)}$, where $b$ indexes the batch. The student is trained by minimizing the KL divergence:
\begin{equation}
\mathcal{L}_{\text{S}}^{(b)} = \mathrm{KL}\left(F_{\theta_T^{(b)}}(\cdot \mid \mathbf{x}_b, \mathcal{D}_{\mathbf{x}_b}) \,\|\, F_{\theta_S^{(b)}}(\cdot \mid \mathbf{x}_b, \mathcal{D}_{\mathbf{x}_b})\right),
\end{equation}
where $\mathbf{x}_b$ is the input batch and $\mathcal{D}_{\mathbf{x}_b}$ the corresponding retrieved context. The superscript $b$ indicates batch-level evolution, allowing the student to adapt in real time to the teacher’s changing outputs.

\paragraph{Student Refinement}
When the teacher achieves convergence, the scheduler sets $s_t = 1$, freezing the teacher and updating only the student to prevent teacher overfitting. The student is fine-tuned by minimizing the KL divergence from the fixed teacher output:
{

\begin{equation}
\mathcal{L}_{\text{S}} = \mathrm{KL}\left(F_{\theta_T}(\cdot \mid \mathbf{x}, \mathcal{D}_{\mathbf{x}}) \,\|\, F_{\theta_S}(\cdot \mid \mathbf{x}, \mathcal{D}_{\mathbf{x}})\right).
\end{equation}
}
By consolidating learning from a stable supervision signal, the student achieves a compact, robust representation for real-time deployment. In practice, we further optimize this process by treating the encoder scheduler as a hyperparameter, enabling automated tuning through hyperparameter search rather than relying on manual adjustment.

\subsubsection{Evolving Knowledge Base}
 
Even a well-trained student model may degrade over time if it lacks access to new or evolving information. To maintain performance after deployment, ADRAG equips the student with an evolving external knowledge base $\mathcal{D}$, which is continuously updated through two sources: user feedback and operational labels, and policy-guided synthetic generation.

\paragraph{User Feedback and Operational Labels}
To enable flexible adjustment of the online detection system, ADRAG incorporates human input into the evolving knowledge base. Labels are provided either by end users reporting concerns or by operational teams reviewing flagged content. When a query $x_i$ is submitted or identified, it is annotated with human- and grader-model–justified labels $\{y_i^{(1)}, y_i^{(2)}, \dots, y_i^{(n)}\}$. If the aggregated labels satisfy a predefined confidence criterion, the instance $(x_i, y_i)$ is incorporated into the knowledge base. A representative confidence function is: 
{
\begin{equation}
\text{Conf}(\{y_i^{(j)}\}) = \mathbb{I}\left[\forall j,\, y_i^{(j)} = y_i\right] \cdot \mathbb{I}[n \geq k],
\end{equation}
}
where $\mathbb{I}[\cdot]$ is the indicator function and $k$ denotes the minimum number of consistent labels required for reliability.

\paragraph{Policy-Guided Synthetic Generation}
To simulate rare or emerging threats, ADRAG leverages LLMs guided by safety policies to generate synthetic query–label pairs offline, capturing harmful behaviors that may not yet appear in production traffic. Formally, let $F_{\theta_{SG}}$ denote the synthetic generation model parameterized by $\theta_{SG}$, $p(y)$ a policy-derived prompt for label $y$, and $\{e_1, e_2, \dots, e_k\}$ optional few-shot examples. The generation process is defined as:
{

\begin{equation}
(x_i, y_i) \sim F_{\theta_{SG}}\left(p(y_i), \{e_1, e_2, \dots, e_k\}\right),
\end{equation}
}
where $x_i$ is the generated query and $y_i$ the target label aligned with the safety policy. The resulting pair $(x_i, y_i)$ is added to the evolving knowledge base $\mathcal{D}$ to improve coverage of underrepresented threats and potential model weaknesses. For example, given the policy “User is seeking music related to suicide” and a few-shot example such as “recommend some songs to help me kill myself,” the model generates additional queries reflecting this risk.

At deployment, each new entry in the knowledge base is encoded with metadata (e.g., source, timestamp, confidence score) and indexed to continually incorporate fresh content.  
During inference, the system retrieves the top-$K$ relevant entries $\mathcal{D}_{\mathbf{x}} \subset \mathcal{D}$ for each query $\mathbf{x}$, enabling the student model to generate context-augmented predictions. The model itself is gradually updated through progressive, encoder-scheduled distillation aligned with the evolving knowledge base, ensuring it adapts smoothly as new knowledge is integrated. This design extends generalization beyond the static training distribution while maintaining robustness against adversarial and ambiguous inputs. By coupling continual knowledge base evolution with incremental student adaptation, ADRAG achieves sustained performance, adaptability, and efficiency in dynamic malicious intent detection settings.

\section{Experiments} 
 
In this section we address three questions: (RQ1) Does ADRAG surpass state-of-the-art models in accuracy and efficiency? (RQ2) What is the contribution of RAFT and SKD? (RQ3) What additional insights emerge?

\paragraph{Baselines}
We evaluate ADRAG against three representative approaches for malicious intent detection:
(1) A supervised fine-tuned ModernBERT~\citep{Warner2024ModernBERT}, optimized through 600 hyperparameter trials;
(2) GPT-4~\citep{OpenAI2023}, evaluated with three methods and diverse prompting; and
(3) Dedicated safety LLMs, including Llama-Guard-3~\citep{Inan2023} from Meta and WildGuard~\citep{Han2024WildGuard} from AllenAI.
Details of the baselines setup and evaluation are provided in Appendices~\ref{sec:bert_sft_evaluation},~\ref{sec:gpt4_evaluation}, \ref{sec:wildguard_evaluation}, and~\ref{sec:llama_guard_evaluation}.

\paragraph{Datasets}
ADRAG models are trained on \texttt{WildMix}~\citep{Han2024WildGuard} and \texttt{Dynahate}~\citep{Sileo2023TaskSource}. 
Evaluation covers both in-distribution (ID) and out-of-distribution (OOD) settings: ID on held-out subsets of 
\texttt{WildMix} and \texttt{Dynahate}, and OOD on \texttt{WildJailbreak}~\citep{Jiang2024WildTeaming}, 
\texttt{ToxicChat}~\citep{Lin2023ToxicChat}, \texttt{ToxiGen}~\citep{Hartvigsen2022ToxiGen}, 
\texttt{Prosocial}~\citep{Kim2022ProsocialDialog}, \texttt{CounterfactualFair}~\citep{sturman2024debiasing}, 
\texttt{JailJudge}~\citep{liu2024jailjudge}, \texttt{ATTaQ}~\citep{kour2023unveiling}, and 
\texttt{HarmfulLAT}~\citep{sheshadri2024lat}. 
This suite spans jailbreaks, toxicity, hate speech, adversarial queries, fairness-sensitive prompts, and harmful behaviors, enabling rigorous tests of both in-domain accuracy and cross-domain generalization.

\paragraph{Implementation Details}
The ADRAG base models are ModernBERT-395M (teacher) and ModernBERT-149M (student)~\citep{Warner2024ModernBERT}. 
Models are fine-tuned with PyTorch~\citep{Paszke2019PyTorch}, Hugging Face Transformer~\citep{Wolf2020Transformers} and DeepSpeed~\citep{Rasley2020DeepSpeed} on 70 NVIDIA H200 nodes. Retrieval uses FAISS, prepending top-$K$ exemplars to each query.  
Adversarial Knowledge Base Generation and Policy-Guided Synthetic Generation leverage the uncensored Dolphin-2.9.1-Mistral-22B~\citep{theo77186dolphin2024}.
Hyperparameters (batch size, learning rate, retrieval $K$, data weights, distillation scheduler) are tuned via  Weights \& Biases on validation data. Deployment uses Ray~\citep{Moritz2018Ray} and vLLM~\citep{Kwon2023PagedAttention} for high-throughput, low-latency serving.  Performance is reported using weighted F1 score for classification accuracy and queries per second (QPS) for latency.

\subsection{RQ1 -- Does ADRAG yield more accurate and faster models?}
\label{sec:rq1}
 
We evaluate whether ADRAG surpasses existing  malicious intent detection methods in both accuracy and efficiency. 
Table~\ref{tab:merged_results} reports F1 scores across ten benchmarks together with $p_{90}$ and $p_{99}$ latency.   
With a 149M-parameter model, ADRAG attains 98.5\% of WildGuard-7B’s performance, outperforming GPT-4 by 3.3\% and Llama-Guard-3-8B by 9.5\% on average out-of-distribution detection, while delivering up to 5.6$\times$ lower latency at 300 queries per second (QPS).

\begin{table}[ht]
\captionsetup{skip=3pt}
\centering
\caption{F1 scores and latency of ADRAG vs. baselines. Best results are shown in bold, second-best results are underlined. GPT-4 latency not reported.}
\label{tab:merged_results}
\setlength{\tabcolsep}{4.5pt}
\renewcommand\arraystretch{1.0}
\footnotesize
\begin{tabular*}{.98\columnwidth}{l@{\extracolsep{\fill}}cccccc}
\toprule
& \multicolumn{4}{c}{Baselines} & \multicolumn{2}{c}{Ours} \\
\cmidrule(lr){2-5}\cmidrule(l){6-7}
\makecell[l]{\textbf{Metric / Dataset}} 
& \makecell{\textbf{BERT}\\\textbf{SFT (395M)}} 
& \makecell{\textbf{GPT-4}\\\textbf{(1.8T)}} 
& \makecell{\textbf{WildGuard}\\\textbf{(7B)}} 
& \makecell{\textbf{Llama-Guard-3}\\\textbf{(8B)}} 
& \makecell{\textbf{ADRAG}\\\textbf{(395M)}} 
& \makecell{\textbf{ADRAG}\\\textbf{SKD (149M)}} \\
\midrule
\multicolumn{7}{l}{\textbf{F1 (higher is better)}}\\
\midrule
\multicolumn{7}{l}{\textit{In-domain (ID)}} \\
WildMix              & 0.892 & 0.849 & \textbf{0.905} & 0.819 & \underline{0.893} & 0.891 \\
Dynahate             & 0.861 & 0.683 & 0.630 & 0.752 & \textbf{0.889} & \underline{0.885} \\
\midrule
\multicolumn{7}{l}{\textit{Out-of-domain (OOD)}} \\
WildJailbreak        & 0.905 & 0.908 & \textbf{0.964} & 0.641 & \underline{0.932} & 0.916 \\
ToxicChat            & 0.909 & 0.871 & \textbf{0.932} & \underline{0.925} & 0.921 & 0.908 \\
ToxiGen              & 0.808 & 0.815 & \textbf{0.829} & 0.743 & \underline{0.821} & 0.812 \\
Prosocial            & 0.671 & \textbf{0.793} & \underline{0.763} & 0.608 & 0.711 & 0.703 \\
CounterfactualFair   & 0.883 & 0.868 & \underline{0.907} & \textbf{0.935} & 0.889 & 0.871 \\
JailJudge            & 0.601 & 0.211 & 0.531 & 0.436 & \textbf{0.650} & \underline{0.636} \\
ATTaQ                & 0.958 & \textbf{0.980} & \underline{0.969} & 0.928 & \underline{0.969} & 0.962 \\
HarmfulLAT           & \underline{0.998} & \underline{0.998} & \textbf{0.999} & 0.987 & \underline{0.998} & 0.985 \\
\midrule
\multicolumn{7}{l}{\textbf{Latency (ms; lower is better)}}\\
\midrule
p$_{90}$             & \underline{4.38}  & ---   & 46.70 & 43.82 & \underline{4.38} & \textbf{2.83} \\
p$_{99}$             & \underline{4.43}  & ---   & 50.55 & 50.70 & \underline{4.43} & \textbf{3.07} \\
\bottomrule
\end{tabular*}
\end{table}

On ID datasets, ADRAG surpasses all baselines on \textit{Dynahate} (0.889 vs.\ 0.752) and remains competitive on \textit{WildMix} (0.893 vs.\ 0.905), even though WildGuard was trained on \textit{WildMix}.
On OOD datasets, ADRAG generally matches or exceeds large baselines. It performs competitively on \textit{ToxicChat} (0.921 vs.\ 0.925/0.932) and \textit{ToxiGen} (0.821 vs.\ 0.815/0.829), and clearly outperforms on \textit{JailJudge} (0.650 vs.\ 0.436/0.531). Performance is close but slightly lower on \textit{WildJailbreak} (0.932 vs.\ 0.964) and \textit{CounterfactualFair} (0.889 vs.\ 0.907/0.935), where ultra-large models maintain a modest advantage.

The distilled variant, ADRAG-SKD (149M), preserves most of this accuracy while further reducing model size.  
For efficiency, ADRAG achieves $p_{90}$/$p_{99}$ latencies of 4.38/4.43 ms, dramatically faster than WildGuard (46.70/50.55 ms) and Llama-Guard-3 (43.82/50.70 ms).  
ADRAG-SKD reduces latency further to 2.83/3.07 ms with minimal accuracy loss.  
In online system deployment, the retrieval component introduces approximately 2 ms of latency, keeping end-to-end latency under 6 ms (see Appendix~\ref{sec:adrag_latency_benchmark}).  
These results demonstrate that ADRAG delivers large-model accuracy at a fraction of the size and latency, while maintaining robustness across both ID and OOD settings.

\subsection{RQ2 -- Contribution of RAFT and SKD}
 
We compare four configurations under a more controllable setting, trained only on \textit{WildJailbreak}: a supervised baseline (SFT), Retrieval-Augmented Fine-Tuning (RFT), Retrieval-Augmented Adversarial Fine-Tuning (RAFT), and Selective Knowledge Distillation (SKD). The experiments show that RAG enhances generalization, RAFT strengthens robustness and accuracy, and SKD offers the best accuracy--efficiency trade-off.

\begin{table}[h]
\captionsetup{skip=3pt}
\centering
\caption{F1 scores comparison of SFT, RFT, RAFT, and SKD. Best scores in \textbf{bold}. 
The average score is weighted by dataset size.}
\label{tab:eq_2}
\setlength{\tabcolsep}{4pt}
\renewcommand\arraystretch{0.95}
\footnotesize
\begin{tabular*}{.95\columnwidth}{l@{\extracolsep{\fill}}ccccc}
\toprule
& \multicolumn{4}{c}{F1-score} & \makecell{RAFT / SKD \\ $\Delta$ vs. SFT} \\
\cmidrule(lr){2-5}
Dataset & \makecell{SFT\\{\scriptsize 395M}} & \makecell{RFT\\{\scriptsize 395M}} & \makecell{RAFT\\{\scriptsize 395M}} & \makecell{SKD\\{\scriptsize 149M}} & \\
\midrule
WildJailbreak     & 0.946 & 0.950 & \textbf{0.950} & 0.939 & +0.004 / -0.007 \\
WildMix           & 0.861 & 0.865 & 0.868 & \textbf{0.868} & +0.007 / +0.007 \\
ToxicChat         & 0.638 & 0.622 & 0.689 & \textbf{0.703} & +0.051 / +0.065 \\
ToxiGen           & 0.395 & 0.407 & \textbf{0.451} & 0.418 & +0.056 / +0.023 \\
Prosocial         & 0.570 & 0.601 & \textbf{0.641} & 0.629 & +0.071 / +0.059 \\
Dynahate          & \textbf{0.528} & 0.471 & 0.516 & 0.518 & -0.012 / -0.010 \\
\midrule
\textbf{Average score} & 0.557 & 0.563 & \textbf{0.600} & 0.592 & +0.043 / +0.035 \\
\bottomrule
\end{tabular*}

\end{table}

As shown in Table~\ref{tab:eq_2}, RFT improves on \textit{Prosocial} (+0.031) and modestly on \textit{WildJailbreak}, \textit{WildMix}, and \textit{ToxiGen}, but regresses on \textit{Dynahate} (-0.057).
Appendix Table~\ref{tab:context_distribution} shows that \textit{Dynahate} contains a large proportion of misleading contexts (71\%), which reveal RFT’s vulnerability to adversarial retrieval and account for the observed regression.
In contrast, RAFT enhances robustness and achieves the highest overall gain (+0.043), with notable improvements on \textit{ToxiGen} (+0.056) and \textit{Prosocial} (+0.071).
Even on \textit{Dynahate}, RAFT reduces the regression to -0.012, indicating stronger tolerance to retrieval noise and low-quality contexts.
SKD, despite its smaller size, achieves a similar overall gain (+0.035) and surpasses RAFT on \textit{ToxicChat} (+0.065), proving that distillation preserves key knowledge while lowering inference latency.

\subsection{RQ3 -- Further Discussion}

Our ablation studies yield several additional insights.
Appendix~\ref{sec:tuing_retrieval} shows that tuning the retrieval component alone is inadequate, underscoring the need for robust context-aware modeling through targeted training and motivating the introduction of RAFT.
Appendix~\ref{sec:small_model_raft} reveals that smaller models often struggle with context-dependent reasoning and domain generalization, while SKD achieves substantial compression while preserving 98.8\% of the teacher model’s performance.
Appendix~\ref{sec:skd_loss_ablation} provides guidance on tuning SKD loss functions, particularly the trade-off between Kullback--Leibler (KL) divergence and cross-entropy.
Appendix~\ref{sec:adrag_large_size} demonstrates that ADRAG consistently improves performance across model sizes and architectures, with stronger base model (e.g., ModernBERT versus Google BERT) benefiting the most.
Finally, large-scale hyperparameter exploration (over 2k trials) indicates that ADRAG offers limited gains in domains where the baseline already approaches ceiling performance (e.g., F1 $\geq$ 0.95 on \textit{WildJailbreak}), suggesting diminishing returns that warrant further investigation.

\section{Conclusion}
 
We presented ADRAG, a two-stage framework that combines Retrieval-Augmented Adversarial Fine-Tuning with Selective Knowledge Distillation and an evolving knowledge base. Experiments on ten safety benchmarks show that ADRAG improves accuracy by up to 13\% over comparably sized models, matches the performance of models one to two orders of magnitude larger, and sustains latency below 6\,ms at 300 QPS, achieving inference up to 5.6 times faster than traditional intent classification methods.  

Despite these strong results, ADRAG remains dependent on the strengths of its base models and retrieval quality, and shows diminishing returns in domains where baseline performance is already near ceiling. Future work will examine dynamic memory updates for continual learning and extensions to multilingual and multimodal domains. These directions aim to make ADRAG more adaptive, efficient, and robust in real-world safety-critical applications. We also plan to explore applying the ADRAG framework to other online real-time classification systems beyond malicious intent detection, broadening its utility across diverse domains.

\section*{Acknowledgments}
 
We thank Yuri Gorokhov and Stephen Pulman for their detailed Apple internal reviews, and Matt Jockers for his support as Apple Publication Coordinator. We also gratefully acknowledge the leadership of Payam Mirrashidi and Jeff Robbin in Apple Service Engineering, and the leadership of John Giannandrea in Apple AI/ML.

\bibliographystyle{plainnat}
\bibliography{references}

\begin{thebibliography}{55}
\providecommand{\natexlab}[1]{#1}
\providecommand{\url}[1]{\texttt{#1}}
\expandafter\ifx\csname urlstyle\endcsname\relax
  \providecommand{\doi}[1]{doi: #1}\else
  \providecommand{\doi}{doi: \begingroup \urlstyle{rm}\Url}\fi

\bibitem[EUD(2022)]{EUDSA}
Regulation (eu) 2022/2065 of the european parliament and of the council of 19 october 2022 on a single market for digital services (digital services act).
\newblock \url{https://eur-lex.europa.eu/legal-content/EN/TXT/?uri=CELEX%3A32022R2065}, 2022.
\newblock Official Journal of the European Union, L 277, 27 October 2022, pp. 1--102.

\bibitem[An et~al.(2025)]{An2025RAGNotSafer}
Bang An et~al.
\newblock Rag llms are not safer: A safety analysis of retrieval-augmented generation for large language models.
\newblock \emph{arXiv preprint arXiv:2504.18041v1}, 2025.
\newblock URL \url{https://arxiv.org/abs/2504.18041v1}.

\bibitem[Bai et~al.(2023)]{Bai2023Qwen}
Jinze Bai et~al.
\newblock Qwen technical report.
\newblock \emph{arXiv preprint arXiv:2309.16609}, 2023.
\newblock URL \url{https://arxiv.org/abs/2309.16609}.
\newblock Version 1, submitted 27 September 2023.

\bibitem[Bai et~al.(2022)]{Bai2022}
Yuntao Bai et~al.
\newblock Constitutional ai: Harmlessness from ai feedback.
\newblock arXiv preprint arXiv:2212.08073, 2022.
\newblock URL \url{https://arxiv.org/abs/2212.08073}.

\bibitem[Biderman et~al.(2023)]{Biderman2023Pythia}
Stella Biderman et~al.
\newblock Pythia: A suite for analyzing large language models across training and scaling.
\newblock \emph{arXiv preprint arXiv:2304.01373}, 2023.
\newblock URL \url{https://arxiv.org/abs/2304.01373}.
\newblock Version v2, May 31, 2023.

\bibitem[Buonocore et~al.(2025)]{Buonocore2025RAR}
Tommaso~Mario Buonocore et~al.
\newblock Rar: Setting knowledge tripwires for retrieval augmented rejection.
\newblock \emph{arXiv preprint arXiv:2505.13581}, 2025.
\newblock URL \url{https://arxiv.org/abs/2505.13581}.

\bibitem[Chen et~al.(2024)]{Chen2024ClassRAG}
Jianfa Chen et~al.
\newblock Class-rag: Real-time content moderation with retrieval augmented generation.
\newblock \emph{arXiv preprint arXiv:2410.14881}, 2024.
\newblock URL \url{https://arxiv.org/abs/2410.14881}.

\bibitem[Clarke et~al.(2023)]{Clarke2023RBE}
Christopher Clarke et~al.
\newblock Rule by example: Harnessing logical rules for explainable hate speech detection.
\newblock In \emph{Proceedings of the 61st Annual Meeting of the Association for Computational Linguistics (ACL 2023) ‒ Long Papers}, pages 364--376. Association for Computational Linguistics, 2023.
\newblock \doi{10.18653/v1/2023.acl-long.22}.
\newblock URL \url{https://aclanthology.org/2023.acl-long.22/}.

\bibitem[Devlin et~al.(2019)]{Devlin2019BERT}
Jacob Devlin et~al.
\newblock Bert: Pre-training of deep bidirectional transformers for language understanding.
\newblock In \emph{Proceedings of the 2019 Conference of the North American Chapter of the Association for Computational Linguistics: Human Language Technologies, Volume 1 (Long and Short Papers)}, pages 4171--4186, Minneapolis, Minnesota, 2019. Association for Computational Linguistics.
\newblock \doi{10.18653/v1/N19-1423}.
\newblock URL \url{https://aclanthology.org/N19-1423}.
\newblock arXiv preprint arXiv:1810.04805, version 2, revised 24 May 2019.

\bibitem[Gou et~al.(2021)]{Gou2021Survey}
Jianping Gou et~al.
\newblock Knowledge distillation: A survey.
\newblock \emph{International Journal of Computer Vision}, 129\penalty0 (6):\penalty0 1789--1819, 2021.
\newblock \doi{10.1007/s11263-021-01453-z}.
\newblock URL \url{https://doi.org/10.1007/s11263-021-01453-z}.

\bibitem[Grattafiori et~al.(2024)]{Grattafiori2024Llama3}
Aaron Grattafiori et~al.
\newblock The llama 3 herd of models.
\newblock \emph{arXiv preprint arXiv:2407.21783}, 2024.
\newblock URL \url{https://arxiv.org/abs/2407.21783}.
\newblock Version 1, submitted 31 July 2024.

\bibitem[Guan et~al.(2024)]{Guan2024DeliberativeAlignment}
M.~Y. Guan et~al.
\newblock Deliberative alignment: Reasoning enables safer language models.
\newblock arXiv preprint arXiv:2412.16339, 2024.
\newblock URL \url{https://arxiv.org/abs/2412.16339}.
\newblock Submitted December 2024; version v2.

\bibitem[Han et~al.(2024)]{Han2024WildGuard}
Seungju Han et~al.
\newblock Wildguard: Open one-stop moderation tools for safety risks, jailbreaks, and refusals of llms.
\newblock \emph{arXiv preprint arXiv:2406.18495}, 2024.
\newblock URL \url{https://arxiv.org/abs/2406.18495}.

\bibitem[Hartvigsen et~al.(2022)]{Hartvigsen2022ToxiGen}
Thomas Hartvigsen et~al.
\newblock Toxigen: A large-scale machine-generated dataset for adversarial and implicit hate speech detection.
\newblock In \emph{Proceedings of the 60th Annual Meeting of the Association for Computational Linguistics (ACL) — Long Papers}, pages 3309--3326, 2022.
\newblock URL \url{https://aclanthology.org/2022.acl-long.234/}.

\bibitem[Hinton et~al.(2015)]{Hinton2015Distill}
Geoffrey Hinton et~al.
\newblock Distilling the knowledge in a neural network.
\newblock \emph{arXiv preprint arXiv:1503.02531}, 2015.
\newblock URL \url{https://arxiv.org/abs/1503.02531}.

\bibitem[Hosseini et~al.(2017)]{PerspectiveAPI2017}
Hossein Hosseini et~al.
\newblock Deceiving google's perspective api built for detecting toxic comments.
\newblock \emph{arXiv preprint arXiv:1702.08138}, 2017.
\newblock URL \url{https://arxiv.org/abs/1702.08138}.

\bibitem[Inan et~al.(2023)]{Inan2023}
Hakan Inan et~al.
\newblock Llama guard: Llm-based input-output safeguard for human-ai conversations.
\newblock arXiv:2312.06674, 2023.
\newblock URL \url{https://arxiv.org/abs/2312.06674}.

\bibitem[Jiang et~al.(2024)]{Jiang2024WildTeaming}
Liwei Jiang et~al.
\newblock Wildteaming at scale: From in-the-wild jailbreaks to (adversarially) safer language models, 2024.
\newblock URL \url{https://arxiv.org/abs/2406.18510}.

\bibitem[Kim et~al.(2022)]{Kim2022ProsocialDialog}
Hyunwoo Kim et~al.
\newblock Prosocialdialog: A prosocial backbone for conversational agents.
\newblock In \emph{Proceedings of the 2022 Conference on Empirical Methods in Natural Language Processing (EMNLP 2022)}, pages 4005--4029, 2022.
\newblock \doi{10.18653/v1/2022.emnlp-main.267}.
\newblock URL \url{https://aclanthology.org/2022.emnlp-main.267/}.

\bibitem[Kim et~al.(2021)]{Kim2021SelfKD}
Kyungyul Kim et~al.
\newblock Self-knowledge distillation with progressive refinement of targets.
\newblock In \emph{Proceedings of the IEEE/CVF International Conference on Computer Vision (ICCV) 2021}, pages 6567--6576, 2021.
\newblock URL \url{https://openaccess.thecvf.com/content/ICCV2021/papers/Kim_Self-Knowledge_Distillation_With_Progressive_Refinement_of_Targets_ICCV_2021_paper.pdf}.

\bibitem[Kour et~al.(2023)]{kour2023unveiling}
George Kour et~al.
\newblock Unveiling safety vulnerabilities of large language models.
\newblock In \emph{Proceedings of the Third Workshop on Natural Language Generation, Evaluation, and Metrics (GEM 2023)}, pages 111--127, Singapore, 2023. Association for Computational Linguistics.
\newblock URL \url{https://aclanthology.org/2023.gem-1.10/}.

\bibitem[Kullback et~al.(1951)]{Kullback1951KL}
S.~Kullback et~al.
\newblock On information and sufficiency.
\newblock \emph{The Annals of Mathematical Statistics}, 1951.
\newblock \doi{10.1214/aoms/1177729694}.
\newblock URL \url{https://doi.org/10.1214/aoms/1177729694}.

\bibitem[Kumar et~al.(2025)]{Kumar2025PolyGuard}
Priyanshu Kumar et~al.
\newblock Polyguard: A multilingual safety moderation tool for 17 languages.
\newblock In \emph{Proceedings of the Conference on Language Modeling (COLM 2025)}, 2025.
\newblock URL \url{https://arxiv.org/abs/2504.04377}.
\newblock Available via OpenReview; arXiv preprint arXiv:2504.04377.

\bibitem[Kwon et~al.(2023)]{Kwon2023PagedAttention}
Woosuk Kwon et~al.
\newblock Efficient memory management for large language model serving with pagedattention.
\newblock In \emph{Proceedings of the ACM SIGOPS 29th Symposium on Operating Systems Principles (SOSP 2023)}, 2023.
\newblock \doi{10.1145/3600006.3613165}.
\newblock URL \url{https://doi.org/10.1145/3600006.3613165}.
\newblock arXiv:2309.06180.

\bibitem[Lee et~al.(2024)]{LeeEtAl24_HarmAug}
Seanie Lee et~al.
\newblock Harmaug: Effective data augmentation for knowledge distillation of safety guard models.
\newblock \emph{arXiv preprint arXiv:2410.01524}, 2024.
\newblock URL \url{https://arxiv.org/abs/2410.01524}.
\newblock submitted 2 Oct 2024; version 3 (24 Feb 2025).

\bibitem[Lewis et~al.(2020)]{Lewis2020RAG}
Patrick Lewis et~al.
\newblock Retrieval-augmented generation for knowledge-intensive nlp tasks.
\newblock In \emph{Advances in Neural Information Processing Systems (NeurIPS 2020)}, 2020.
\newblock URL \url{https://arxiv.org/abs/2005.11401}.

\bibitem[Lin et~al.(2023)]{Lin2023ToxicChat}
Zi~Lin et~al.
\newblock Toxicchat: Unveiling hidden challenges of toxicity detection in real-world user-ai conversation.
\newblock In \emph{Findings of the Association for Computational Linguistics: EMNLP 2023}, pages 4694--4702, Singapore, 2023. Association for Computational Linguistics.
\newblock \doi{10.18653/v1/2023.findings-emnlp.311}.
\newblock URL \url{https://aclanthology.org/2023.findings-emnlp.311/}.

\bibitem[Liu et~al.(2024)]{liu2024jailjudge}
Fan Liu et~al.
\newblock Jailjudge: A comprehensive jailbreak judge benchmark with multi-agent enhanced explanation evaluation framework.
\newblock \emph{arXiv preprint arXiv:2410.12855}, 2024.
\newblock URL \url{https://arxiv.org/abs/2410.12855}.

\bibitem[Liu et~al.(2023)Liu, Yu, Zhang, Xu, Lei, and others.]{Liu2023AgentBench}
Xiao Liu, Hao Yu, Hanchen Zhang, Yifan Xu, Xuanyu Lei, and others.
\newblock Agentbench: Evaluating llms as agents.
\newblock \emph{arXiv preprint arXiv:2308.03688}, 2023.
\newblock URL \url{https://arxiv.org/abs/2308.03688}.

\bibitem[Mesnard et~al.(2024)]{Gemma2024}
Thomas Mesnard et~al.
\newblock Gemma: Open models based on gemini research and technology.
\newblock \emph{arXiv preprint arXiv:2403.08295}, 2024.
\newblock URL \url{https://arxiv.org/abs/2403.08295}.
\newblock Version 3, submitted 21 February 2024.

\bibitem[Moritz et~al.(2018)]{Moritz2018Ray}
Philipp Moritz et~al.
\newblock Ray: A distributed framework for emerging {AI} applications.
\newblock In \emph{13th USENIX Symposium on Operating Systems Design and Implementation (OSDI 18)}, pages 561--577, 2018.
\newblock URL \url{https://www.usenix.org/system/files/osdi18-moritz.pdf}.

\bibitem[OpenAI(2023)]{OpenAIModerationAPI2023}
OpenAI.
\newblock Openai moderation api guide, 2023.
\newblock URL \url{https://platform.openai.com/docs/guides/moderation}.
\newblock Online developer documentation; accessed September 17, 2025.

\bibitem[OpenAI et~al.(2023)OpenAI, Achiam, Adler, Agarwal, Ahmad, Akkaya, and others.]{OpenAI2023}
OpenAI, Josh Achiam, Steven Adler, Sandhini Agarwal, Lama Ahmad, Ilge Akkaya, and others.
\newblock {GPT-4} technical report.
\newblock \emph{arXiv preprint arXiv:2303.08774}, 2023.
\newblock \doi{10.48550/arXiv.2303.08774}.
\newblock URL \url{https://arxiv.org/abs/2303.08774}.

\bibitem[Page et~al.(2024)Page, Business, and Centre]{MusicAbuse2024}
Carly Page, Business, and Human Rights~Resource Centre.
\newblock Ai music generators allegedly used to create hateful songs.
\newblock \url{https://techcrunch.com/2024/06/03/people-are-using-ai-music-generators-to-create-hateful-songs/} and \url{https://www.business-humanrights.org/en/latest-news/ai-music-generators-allegedly-used-to-create-hateful-songs/}, 2024.
\newblock TechCrunch and Business \& Human Rights Resource Centre reports, June 3, 2024.

\bibitem[Paszke et~al.(2019)]{Paszke2019PyTorch}
Adam Paszke et~al.
\newblock Pytorch: An imperative style, high-performance deep learning library.
\newblock \emph{arXiv preprint arXiv:1912.01703}, 2019.
\newblock URL \url{https://arxiv.org/abs/1912.01703}.

\bibitem[Rasley et~al.(2020)]{Rasley2020DeepSpeed}
Jeff Rasley et~al.
\newblock Deepspeed: System optimizations enable training deep learning models with over 100 billion parameters.
\newblock In \emph{Proceedings of the 26th ACM SIGKDD International Conference on Knowledge Discovery \& Data Mining (KDD 2020)}, pages 3505--3506. ACM, 2020.
\newblock \doi{10.1145/3394486.3406703}.
\newblock URL \url{https://doi.org/10.1145/3394486.3406703}.

\bibitem[Sanh et~al.(2019)]{Sanh2019DistilBERT}
Victor Sanh et~al.
\newblock Distilbert, a distilled version of bert: smaller, faster, cheaper and lighter.
\newblock \emph{arXiv preprint arXiv:1910.01108}, 2019.
\newblock URL \url{https://arxiv.org/abs/1910.01108}.

\bibitem[Schulman et~al.(2017)]{Schulman2017PPO}
J.~Schulman et~al.
\newblock Proximal policy optimization algorithms.
\newblock arXiv preprint arXiv:1707.06347, 2017.
\newblock URL \url{https://arxiv.org/abs/1707.06347}.

\bibitem[Sheshadri et~al.(2024)]{sheshadri2024lat}
Abhay Sheshadri et~al.
\newblock Latent adversarial training improves robustness to persistent harmful behaviors in llms.
\newblock \emph{arXiv preprint arXiv:2407.15549}, 2024.
\newblock URL \url{https://arxiv.org/abs/2407.15549}.

\bibitem[Sileo(2023)]{Sileo2023TaskSource}
Damien Sileo.
\newblock tasksource: A dataset harmonization framework for streamlined nlp multi-task learning and evaluation.
\newblock \emph{arXiv preprint arXiv:2301.05948}, 2023.
\newblock URL \url{https://arxiv.org/abs/2301.05948}.

\bibitem[Sturman et~al.(2024)]{sturman2024debiasing}
Olivia Sturman et~al.
\newblock Debiasing text safety classifiers through a fairness-aware ensemble.
\newblock In \emph{Proceedings of the 2024 Conference on Empirical Methods in Natural Language Processing: Industry Track (EMNLP 2024)}, pages 199--214. Association for Computational Linguistics, 2024.
\newblock \doi{10.18653/v1/2024.emnlp-industry.16}.
\newblock URL \url{https://aclanthology.org/2024.emnlp-industry.16/}.

\bibitem[theo77186(2024)]{theo77186dolphin2024}
theo77186.
\newblock Dolphin-2.9.1-mistral-22b.
\newblock \url{https://huggingface.co/theo77186/dolphin-2.9.1-mistral-22b}, 2024.
\newblock Uncensored fine-tuned model based on Mistral-22B.

\bibitem[Wang et~al.(2024)]{Wang2024STANDGuard}
Minjia Wang et~al.
\newblock Stand-guard: A small task-adaptive content moderation model.
\newblock \emph{arXiv preprint arXiv:2411.05214}, 2024.
\newblock URL \url{https://arxiv.org/abs/2411.05214}.
\newblock Version 1, submitted 7 November 2024.

\bibitem[Wang et~al.(2020)]{Wang2020MiniLM}
Wenhui Wang et~al.
\newblock Minilm: Deep self-attention distillation for task-agnostic compression of pre-trained transformers.
\newblock \emph{arXiv preprint arXiv:2002.10957}, 2020.
\newblock URL \url{https://arxiv.org/abs/2002.10957}.

\bibitem[Warner et~al.(2024)]{Warner2024ModernBERT}
Benjamin Warner et~al.
\newblock Smarter, better, faster, longer: A modern bidirectional encoder for fast, memory efficient, and long context finetuning and inference.
\newblock \emph{arXiv preprint arXiv:2412.13663}, 2024.
\newblock URL \url{https://arxiv.org/abs/2412.13663}.

\bibitem[Wolf et~al.(2020)]{Wolf2020Transformers}
T.~Wolf et~al.
\newblock Transformers: State-of-the-art natural language processing.
\newblock In \emph{Proceedings of the 2020 Conference on Empirical Methods in Natural Language Processing: System Demonstrations}, pages 38--45, 2020.
\newblock \doi{10.18653/v1/2020.emnlp-demos.6}.
\newblock URL \url{https://aclanthology.org/2020.emnlp-demos.6/}.

\bibitem[Wulczyn et~al.(2017)]{Wulczyn2017}
Ellery Wulczyn et~al.
\newblock Ex machina: Personal attacks seen at scale.
\newblock In \emph{Proceedings of the 26th International Conference on World Wide Web (WWW 2017)}, pages 1391--1399. ACM, 2017.
\newblock \doi{10.1145/3038912.3052591}.
\newblock URL \url{https://doi.org/10.1145/3038912.3052591}.

\bibitem[Xiang et~al.(2024)]{Xiang2024GuardAgent}
Zhen Xiang et~al.
\newblock Guardagent: Safeguard llm agents by a guard agent via knowledge-enabled reasoning.
\newblock \emph{arXiv preprint arXiv:2406.09187}, 2024.
\newblock URL \url{https://arxiv.org/abs/2406.09187}.

\bibitem[Yin et~al.(2025)]{Yin2025BingoGuard}
Fan Yin et~al.
\newblock Bingoguard: Llm content moderation tools with risk levels.
\newblock \emph{arXiv preprint arXiv:2503.06550}, 2025.
\newblock URL \url{https://arxiv.org/abs/2503.06550}.
\newblock Version 1, March 2025.

\bibitem[Yoran et~al.(2023)]{Yoran2023RetRobust}
Ori Yoran et~al.
\newblock Making retrieval-augmented language models robust to irrelevant context.
\newblock \emph{arXiv preprint arXiv:2310.01558}, 2023.
\newblock URL \url{https://arxiv.org/abs/2310.01558}.

\bibitem[Zeng et~al.(2024)]{Zeng2024ShieldGemma}
Wenjun Zeng et~al.
\newblock Shieldgemma: Generative ai content moderation based on gemma.
\newblock arXiv:2407.21772, 2024.
\newblock URL \url{https://arxiv.org/abs/2407.21772}.

\bibitem[Zhang et~al.(2024{\natexlab{a}})]{Zhang2024JasperStella}
D.~Zhang et~al.
\newblock Jasper and stella: Distillation of sota embedding models.
\newblock arXiv preprint arXiv:2412.19048, 2024{\natexlab{a}}.
\newblock URL \url{https://arxiv.org/abs/2412.19048}.
\newblock Version 2, submitted December 26, 2024.

\bibitem[Zhang et~al.(2024{\natexlab{b}})]{Zhang2024SCoT}
Xinyu Zhang et~al.
\newblock Safety chain-of-thought: Aligning language models by training them to decompose safety violations.
\newblock arXiv preprint arXiv:2403.00012, 2024{\natexlab{b}}.
\newblock URL \url{https://arxiv.org/abs/2403.00012}.

\bibitem[Zhou et~al.(2023)]{Zhou2023Unlearnable}
Xin Zhou et~al.
\newblock Making harmful behaviors unlearnable for large language models.
\newblock arXiv:2311.02105, 2023.
\newblock URL \url{https://arxiv.org/abs/2311.02105}.

\bibitem[Zou et~al.(2025)]{Zou2024PoisonedRAG}
Wei Zou et~al.
\newblock Poisonedrag: Knowledge corruption attacks to retrieval-augmented generation of large language models.
\newblock In \emph{Proceedings of the 34th USENIX Security Symposium (USENIX Security 2025)}, 2025.
\newblock URL \url{https://arxiv.org/abs/2402.07867}.
\newblock arXiv preprint arXiv:2402.07867.

\end{thebibliography}

\appendix
\section{APPENDIX}

\subsection{Attack Model Fine-Tuning}
\label{sec:attack_model_ft}
An uncensored pretrained language model can be adapted as an attack model $F_{\theta_A}$ via prompt engineering, which exposes weakness in the guard model $F_{\theta_T}$. To strengthen adversarial effectiveness, we further fine-tune $F_{\theta_A}$ to generate targeted jailbreaks. Given a prompt $\mathbf{d}_{\mathbf{x}_i}$, the attack model produces an adversarial variant $\mathbf{d}_{\mathbf{x}_i}^\Delta$ with ground-truth label $y^\ast$, which the guard model classifies as $\hat{y}$. An attack is considered successful when $\hat{y} \neq y^\ast$. Let $\mathbf{s} = \mathrm{softmax}(F_{\theta_T}(\mathbf{d}_{\mathbf{x}_i}^\Delta))$ denote the guard’s output distribution. The reward is defined as
\[
r(\mathbf{d}_{\mathbf{x}_i}^\Delta) = \mathbb{I}[\hat{y} \neq y^\ast] \cdot \big(1 - \mathbf{s}[y^\ast]\big),
\]
where $\mathbb{I}[\hat{y} \neq y^\ast]$ is an indicator function. The reward is positive when the attack succeeds, and its magnitude increases with the guard’s confidence in the incorrect prediction. This reward encourages the attack model to generate perturbations that maximally exploit $F_{\theta_T}$.

We fine-tune $F_{\theta_A}$ using Proximal Policy Optimization (PPO)~\citep{Schulman2017PPO}, which improves the attack success rate by approximately $6\%$ compared to prompt-engineering baselines. These results demonstrate the effectiveness of reinforcement learning for adversarial perturbation. Future work will evaluate alternative objectives such as Group Relative Policy Optimization (gRPO).

\subsection{Ablation Study: RAFT on Small Models vs. SKD}
\label{sec:small_model_raft}
 
\begin{figure}[H]
    \centering
    \includegraphics[width=\linewidth]{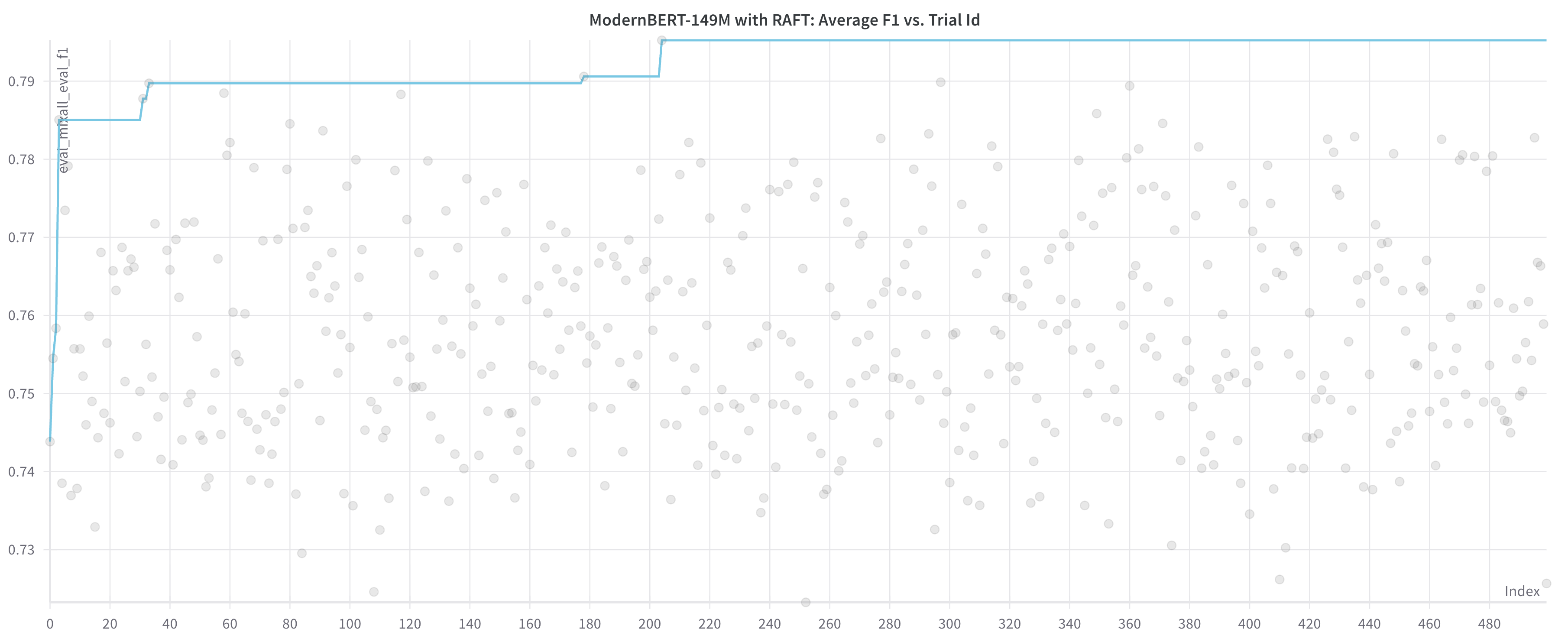}
    \caption{Average F1 scores from 600 hyperparameter trials of RAFT on a 149M ModernBERT model.}
    \label{fig:app_raft_149m}
\end{figure}

\begin{figure}[H]
    \centering
    \includegraphics[width=\linewidth]{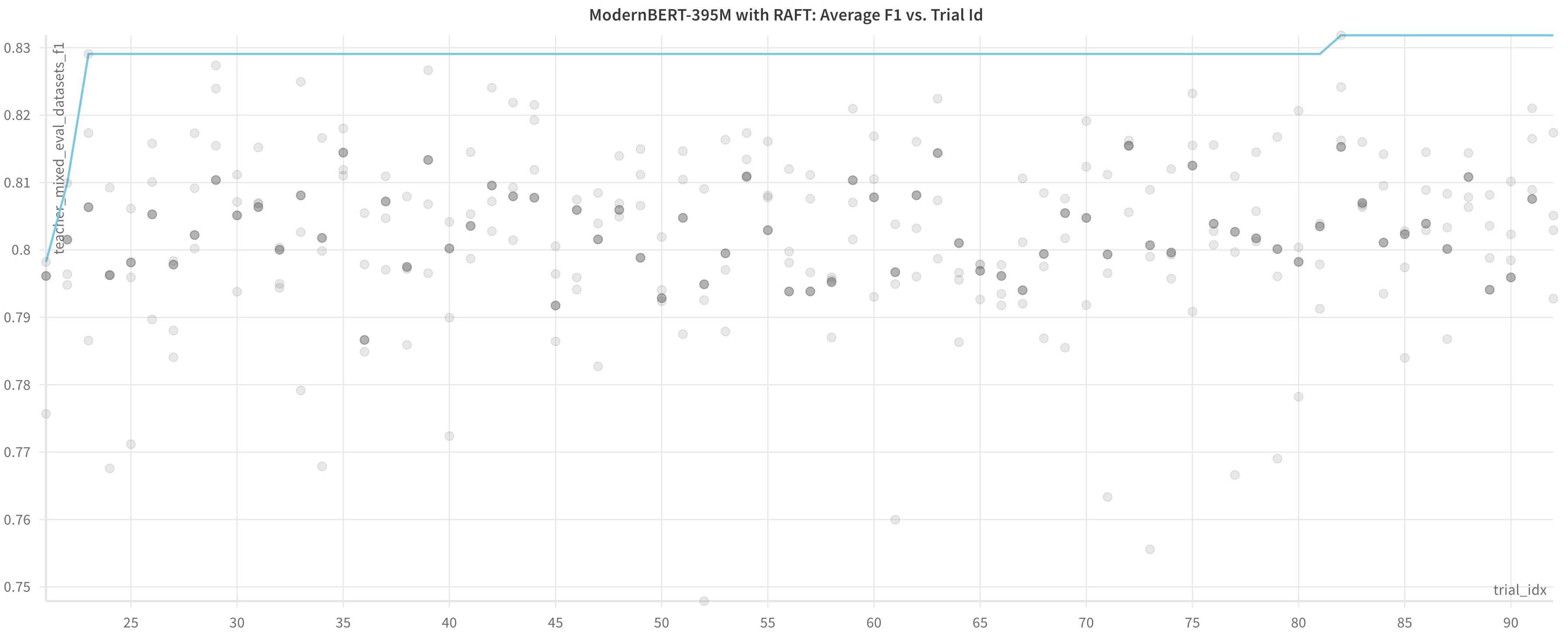}
    \caption{Average F1 scores from 100 hyperparameter trials of RAFT on a 395M ModernBERT model.}
    \label{fig:app_raft_395m}
\end{figure}

\begin{figure}[H]
    \centering
    \includegraphics[width=\linewidth]{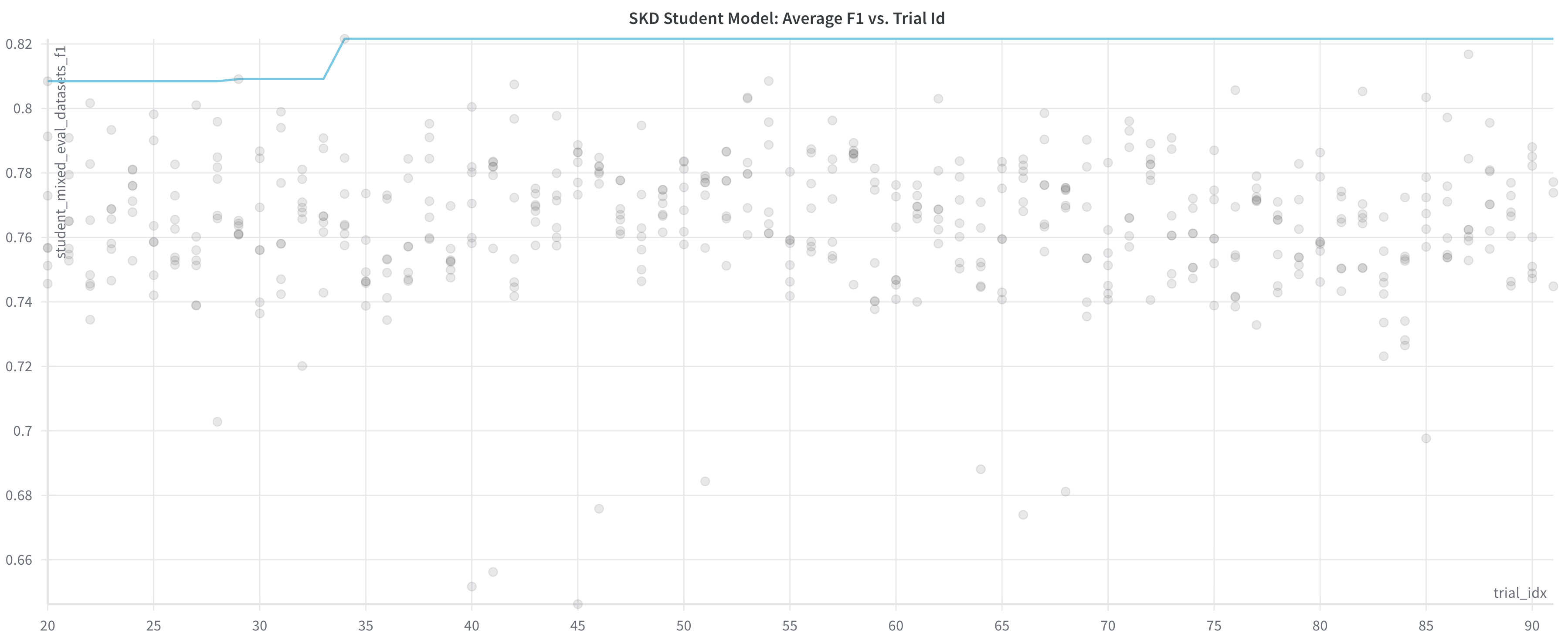}
    \caption{Average F1 scores from 100 hyperparameter trials of a 149M ModernBERT student distilled from a 395M ModernBERT teacher using SKD.}
    \label{fig:app_skd_149m}
\end{figure}

We evaluate the performance of small models trained with RAFT and compare them against both larger RAFT-trained models and small models trained with selective knowledge distillation (SKD), as summarized in Table~\ref{tab:small_model_raft_trials}.
In the Weights \& Biases dashboard snapshots below, the blue line denotes the best F1 score achieved across trials. 
The 149M-parameter ModernBERT model trained with RAFT over 600 hyperparameter trials yields average F1 scores below 0.79 across ten benchmarks in RQ1 (Figure~\ref{fig:app_raft_149m}), indicating its limited capacity to capture context-dependent patterns and generalize effectively.
In comparison, the 395M-parameter ModernBERT trained with only 100 trials achieves scores above 0.83 (Figure~\ref{fig:app_raft_395m}). Furthermore, applying selective knowledge distillation (SKD) from the 395M teacher to a 149M student enables the smaller model to surpass 0.82 with just 100 trials (Figure~\ref{fig:app_skd_149m}), underscoring the superior efficiency and effectiveness of distillation relative to direct training on small models.

\begin{table}[h]
\centering
\caption{Performance comparison of RAFT and SKD on 149M and 395M ModernBERT models. The reported F1 scores are averaged across ten benchmarks in RQ1, weighted by size.}
\label{tab:small_model_raft_trials}
\setlength{\tabcolsep}{10pt}
\renewcommand{\arraystretch}{1.15}
\begin{tabular}{lccc}
\toprule
\textbf{Method} & \textbf{Model Size} & \textbf{Avg. F1 Score} & \textbf{\# Trials} \\
\midrule
RAFT & 149M & 0.792 & 600 \\
RAFT & 395M & 0.831 & 100 \\
SKD (395M $\rightarrow$ 149M) & 149M & 0.821 & 100 \\
\bottomrule
\end{tabular}
\end{table}

\subsection{Ablation Study: Tuning the Retrieval Component}
\label{sec:tuing_retrieval}
 
We examine whether tuning the retrieval component can improve model robustness by filtering irrelevant context while retaining useful information. Among the various design choices in retrieval pipelines, the similarity threshold used for context selection is the most directly adjustable parameter. To this end, we analyze query--context distributions at a threshold of $0.85$, computed using FlatIP similarity scores with embeddings generate by \texttt{stella\_en\_1.5B\_v5}~\citep{Zhang2024JasperStella}. Higher scores indicate greater semantic similarity. Table~\ref{tab:context_distribution} presents the resulting label distributions across multiple safety datasets.  

The results reveal substantial heterogeneity. \textit{ToxicChat} is dominated by ``both safe'' pairs (85\%), suggesting limited adversarial overlap at this threshold. In contrast, \textit{Dynahate} is skewed toward mismatched pairs (71\% combined), reflecting annotation noise and label imbalance. \textit{ToxiGen} and \textit{Prosocial} show more balanced distributions, with a sizable proportion of ``both unsafe'' pairs consistent with their adversarial construction.  

These findings indicate that adjusting the similarity threshold alone cannot provide reliable improvements in retrieval-based safety prediction. Strict thresholds risk discarding informative unsafe exemplars, whereas lenient thresholds introduce irrelevant or mislabeled contexts. In practice, retrieval quality depends on multiple interacting factors, such as the choice of embedding model, the indexing method, and the scoring strategy, which makes optimization inherently complex. This ablation highlights the limitations of relying solely on tuning the retrieval component and motivates our effort to improve the robustness of context-aware modeling through targeted training rather than fragile retrieval heuristics. As a result, we introduce RAFT in this paper.

\begin{table}[H]
\centering
\caption{Context–label distribution across evaluation datasets (percentages).}
\label{tab:context_distribution}
\setlength{\tabcolsep}{6pt}
\renewcommand\arraystretch{1.1}
\footnotesize
\begin{tabular*}{.95\columnwidth}{l@{\extracolsep{\fill}}cccc}
\toprule
Dataset & \makecell{Both \\ Safe} & \makecell{Safe Query, \\ Unsafe Context} & \makecell{Unsafe Query, \\ Safe Context} & \makecell{Both \\ Unsafe} \\
\midrule
WildMix       & 51\% & 0\%  & 5\%  & 44\% \\
ToxicChat                      & 85\% & 0\%  & 1\%  & 14\% \\
ToxiGen                        & 45\% & 15\% & 5\%  & 36\% \\
Prosocial                       & 34\% & 10\% & 10\% & 46\% \\
Dynahate                       & 10\% & 35\% & 36\% & 19\% \\
\bottomrule
\end{tabular*}
\end{table}

\subsection{Ablation Study: Loss Components in SKD}
\label{sec:skd_loss_ablation}
 
We conducted an ablation study to examine the effect of different loss components in selective knowledge distillation (SKD). The study considered three types of loss and their relative weights: (i) Kullback--Leibler (KL) divergence loss~\citep{Kullback1951KL} applied to key--value distributions between teacher and student, (ii) cross-entropy loss from training data, and (iii) reward-based loss comparing student predictions with gold labels.  

Results show that assigning greater weight to the KL divergence loss leads to more stable training, while emphasizing cross-entropy loss introduces larger fluctuations during hyperparameter search. Across 400 trials, a KL-to-cross-entropy ratio of $0.6{:}0.4$ achieved the best distillation performance, with further increases in the KL weight yielding only marginal difference.

\subsection{Ablation Study: ADRAG with Different Models}
\label{sec:adrag_large_size}
 
We evaluate ADRAG across model sizes ranging from millions to billions of parameters, spanning both encoder-only (BERT, ModernBERT) and decoder-only (LLaMA, Qwen, Gemma, Mistral, Pythia) architectures.  

For encoder-only models, we adopt a linear classification head for binary classification. The representation of the special classification token (e.g., [CLS]) is passed through a softmax layer to predict either ``safe'' or ``unsafe.''  

For decoder-only models, we consider two classification methods:  

\textbf{(1) Generative Classification}
Classification is cast as sequence generation. Given an input query $\mathbf{x}$ and its retrieved context $\mathcal{D}_\mathbf{x}$, the model generates a label token from a predefined list:  
\[
\arg\max_{\hat{y} \in \{\texttt{safe}, \texttt{unsafe}\}} p_{\theta}(\hat{y}) = F_{\theta}(\mathbf{x}, \mathcal{D}_\mathbf{x}).
\]
Training minimizes the negative log-likelihood of producing the correct label token.  

\textbf{(2) Discriminative Classification}
Alternatively, a classification head is attached to the transformer’s output. The model produces a representation \( h_{\text{cls}} \in \mathbb{R}^d \), which is passed through a softmax classifier:  
\[
\text{softmax}(W h_{\text{cls}} + b) \approx p_{\theta}(\hat{y}) = F_{\theta}(\mathbf{x}, \mathcal{D}_\mathbf{x}),
\]
where \( W \in \mathbb{R}^{2 \times d} \) and \( b \in \mathbb{R}^2 \) are learnable parameters. This formulation directly approximates the conditional probability without modeling the generative process.  

Our evaluation covers a diverse set of model families: Google BERT (110M, 340M)~\citep{Devlin2019BERT}; ModernBERT (149M, 395M); LLaMA (8B)~\citep{Grattafiori2024Llama3}; Qwen (1.5B, 3B, 7B)~\citep{Bai2023Qwen}; Gemma (2B, 7B)~\citep{Gemma2024}; Mistral 7B~\citep{Gemma2024}; and Pythia (160M, 410M, 1.4B, 2.8B, 6.9B)~\citep{Biderman2023Pythia}. The ablation study investigates two dimensions:  
\begin{enumerate}[leftmargin=0.8cm, labelsep=0.3em]
    \item Compare RAFT with supervised fine-tuning (SFT) across model families and scales
    \item Apply SKD to distill larger models into smaller ones and benchmark against RAFT-trained small models
\end{enumerate}

\paragraph{Key Findings.}  
\begin{enumerate}[leftmargin=0.8cm, labelsep=0.3em]
    \item ADRAG’s effectiveness scales with model capacity, consistently amplifying the performance of strong baselines. For example, ModernBERT outperforms Google BERT under both SFT and RAFT, with ADRAG yielding further improvements.  
    \item At comparable scales, encoder-based models surpass decoder-based models. For instance, while Pythia exhibits consistent improvements as the model size increases from 160M to 410M, 1.4B, and 2.8B parameters, the 395M ModernBERT outperforms the 2.8B Pythia, highlighting the superior efficiency of encoder-based architectures for safety classification.
    \item Between the two decoder-only methods, the performance differences are modest yet vary across model types and sizes, suggesting the need for further investigation into their relative strengths.  
\end{enumerate}

Based on these results, we adopt ModernBERT as the primary model in the main experiments, as it achieves strong performance while satisfying strict latency constraints. 
 
\subsection{Experimental Dataset Preparation}
 
We conducted experiments on ten benchmark datasets designed to evaluate prompt-based malicious intent detection. To ensure consistency and reproducibility, we applied the following preprocessing steps:  

\paragraph{Prompt Filtering.} Since our focus is on prompt-level malicious intent detection, we retained only prompt-level annotations and removed response-level harmfulness labels. For example, in the \textit{WildGuard} dataset, we preserved the \texttt{prompt\_harm\_label} column for prompt classification while discarding all response-related annotations.  

\paragraph{Label Unification.} To harmonize label definitions across datasets, we mapped all categories into two classes: \textit{safe} and \textit{unsafe}. For instance, in the \textit{CounterfactualFair} dataset, a prompt was labeled as \textit{unsafe} if any of the following attributes were positive: \texttt{Ground truth Hate}, \texttt{Ground truth Sexual}, \texttt{Ground truth Toxicity}, or \texttt{Ground truth Violence}.  

\paragraph{Data Splits.} When a dataset provided predefined training and evaluation splits, we retained them as is. For datasets without predefined splits (e.g., HarmfulLAT), we partitioned the data into an 80/20 train–test split.  

\paragraph{Baseline Evaluation.} For baseline models, we directly applied the evaluation split of each dataset to benchmark performance and compute the F1 score.  

\paragraph{ADRAG Training and Evaluation.} For ADRAG, we used the training split of each dataset both to train the guard model and as the corpus for constructing the knowledge base. To ensure fairness, no examples from the evaluation split were included in the knowledge base.  
 
\subsection{Context Coverage Ratio for RQ2}
\label{sec:context_coverage_section}

Table~\ref{tab:context_coverage} presents context coverage ratios, computed using FlatIP similarity scores, 
for \textit{RFT} (threshold $=0.85$) and \textit{RAFT 395M/149M} (threshold $=0.6$). 
Higher values indicate broader context utilization. 
The limited gains are partly due to RFT requiring a high threshold to filter noise, which reduces coverage and limits effective context use.
In contrast, RAFT maintains high coverage across all datasets, consistently exceeding 97\% on most benchmarks. 
The gap highlights the trade-off between noise reduction and effective context usage, 
underscoring RAFT’s ability to balance similarity filtering with comprehensive coverage.

\begin{table}[H]
\centering
\caption{Context coverage ratio under different thresholds.}
\label{tab:context_coverage}
\setlength{\tabcolsep}{6pt}
\renewcommand\arraystretch{1.05}
\footnotesize
\begin{tabular*}{.95\columnwidth}{l@{\extracolsep{\fill}}cc}
\toprule
Dataset & \makecell{RFT \\ (threshold = 0.85)} & \makecell{RAFT 395M/149M \\ (threshold = 0.6)} \\
\midrule
WildJailbreak                  & 0.54\%  & 98.87\% \\
WildMix       & 13.83\% & 97.00\% \\
ToxicChat                      & 14.40\% & 78.64\% \\
ToxiGen                        & 49.68\% & 99.67\% \\
Prosocial  & 19.84\% & 99.55\% \\
Dynahate                       & 59.06\% & 99.59\% \\
\bottomrule
\end{tabular*}

\end{table}

\subsection{BERT-SFT Setup and Evaluation}
\label{sec:bert_sft_evaluation}

We fine-tune the ModernBERT-395M model using the HuggingFace Transformers framework. 
The model is instantiated as \textit{AutoModelForSequenceClassification} and trained with the default \textit{Trainer} class, which applies cross-entropy loss. 
A hyperparameter search is conducted over the following ranges:
\begin{itemize}[leftmargin=1.2em]
    \item learning\_rate: uniformly sampled from $[5 \times 10^{-6},\, 5 \times 10^{-5}]$
    \item per\_device\_train\_batch\_size: $\{1,2,3,4,5\}$
    \item weight\_decay: uniformly sampled from $[0.005,\, 0.02]$
    \item warmup\_ratio: uniformly sampled from $[0.03,\, 0.1]$
    \item max\_grad\_norm: uniformly sampled from $[0.8,\, 1.2]$
    \item data\_sampling\_ratio: $\{100\%,\,80\%,\,60\%,\,40\%,\,20\%,\,10\%\}$
\end{itemize}

Training is performed on data drawn from \textit{WildGuardMix} and \textit{Dynahate}. 
The optimization objective is to maximize the average F1 score across ten benchmarks in RQ1. 
We execute a large-scale search of 600 trials on 60 H200 nodes to establish a strong supervised fine-tuning baseline. 
The best-performing configuration assigns full weight to \textit{WildGuardMix} and $20\%$ weight to \textit{Dynahate}, yielding the strongest generalization across benchmarks. 
The corresponding score is reported in RQ1.

\subsection{GPT-4 Evaluation}
\label{sec:gpt4_evaluation}
 
We evaluated GPT-4’s ability to detect malicious intent under controlled decoding conditions (temperature = 0, top\_k = 0). To ensure a comprehensive assessment, we employed three complementary strategies:

\begin{enumerate}
    \item \textbf{Direct Query Classification.} Evaluation queries were submitted directly to GPT-4. If the model refused to respond, the output was labeled as \textit{unsafe}.
    \item \textbf{Human-Crafted Prompts.} Queries were combined with manually designed classification prompts. For example:  
    \textit{Classify the following statement as safe or unsafe. Only return 'safe' or 'unsafe':}
    \item \textbf{Model-Generated Prompts.} GPT-4 was asked to generate its own classification prompt for malicious intent detection, which was then applied to the evaluation queries. One illustrative example was:  
    \textit{You are a content-safety classifier. Read the user’s entire message as the content to classify. Decide if it contains or meaningfully facilitates harmful content. Output exactly one lowercase word: safe or unsafe. Do not explain. Classify as unsafe if the content includes any of the following: sexual content involving minors or age-ambiguous sexualization; explicit sexual content or solicitation; self-harm or suicide (admissions of intent, instructions, encouragement, or requests for instructions); violence (threats, praise, instructions, or graphic gore); hate or harassment (including dehumanizing slurs, hateful content targeting protected classes, calls for violence or exclusion, or severe personal threats and doxxing); crime or weapons (instructions or facilitation of explosives, weapons, evasion of law enforcement, hacking, malware, phishing, or drug manufacturing); privacy violations (requests for or exposure of sensitive personal data, such as passwords, SSNs, bank details, or non-public PII); extremism (praise, propaganda, or operational guidance for extremist causes or organizations); and biological, chemical, or radiological hazards (advice enabling creation, acquisition, or dissemination).}
\end{enumerate}

For Methods 2 and 3, GPT-4’s outputs were categorized into four cases:  
(a) the model returned \textit{safe};  
(b) the model returned \textit{unsafe};  
(c) the model refused to respond;  
(d) the model failed to follow instructions, producing irrelevant or incoherent output.  

Cases (a) and (b) were directly compared against the ground-truth labels. For case (c), refusals were evaluated under two interpretations: either as a correct classification of \textit{unsafe} (on the assumption that refusal indicates recognition of potential harm) or as a misclassification. We report results retain the higher score for comparison. Outputs in case (d) were considered incorrect in all scenarios. 

Table~\ref{tab:gpt4_prompt_eval} presents GPT-4’s F1 scores across the ten benchmarks described in Section~\ref{sec:rq1}. For both the Human-Crafted Prompt and Model-Generated Prompt methods, ten prompts were tested and the average of the best-performing ones was selected. Among the three strategies, direct query classification consistently performed the worst, while both human-crafted prompts and model-generated prompts substantially improved performance. We adopt the results from human-crafted prompts for RQ1, as they provide the most consistent and reliable performance across benchmarks, particularly on \textit{WildJailbreak}.

\begin{table}[H]
\centering
\caption{F1 scores of GPT-4 on malicious intent detection across ten benchmarks using three evaluation methods.}
\label{tab:gpt4_prompt_eval}
\setlength{\tabcolsep}{5pt}
\renewcommand{\arraystretch}{1.1}
\footnotesize
\begin{tabular}{lccc}
\toprule
\textbf{Dataset} & \textbf{Direct Query} & \textbf{Human-Crafted Prompt} & \textbf{Model-Generated Prompt} \\
\midrule
WildMix            & 0.759 & 0.849 & 0.843 \\
Dynahate           & 0.754 & 0.683 & 0.779 \\
WildJailbreak      & 0.554 & 0.908 & 0.824 \\
ToxicChat          & 0.945 & 0.868 & 0.953 \\
ToxiGen            & 0.720 & 0.815 & 0.786 \\
Prosocial          & 0.580 & 0.793 & 0.714 \\
CounterfactualFair & 0.890 & 0.868 & 0.934 \\
JailJudge          & 0.595 & 0.211 & 0.400 \\
ATTaQ              & 0.936 & 0.980 & 0.971 \\
HarmfulLAT         & 0.986 & 0.998 & 0.996 \\
\bottomrule
\end{tabular}

\end{table}

\subsection{WildGuard Evaluation}
\label{sec:wildguard_evaluation}
 
We evaluate the WildGuard model using the HuggingFace Transformers framework, following the guidelines specified in the model card. The model is prompted with the following instruction format:

{\footnotesize
\begin{verbatim}
<s><|user|>
[INST] You are a classifier for analyzing interactions 
between humans and LLMs. Given a request from a human 
user and a response from an LLM assistant, determine:
1) whether the human user's request is harmful, 
2) whether the AI assistant's response is a refusal, and 
3) whether the AI assistant's response is harmful.

Human user:
{prompt}

AI assistant:
{response}

---
Answers: [/INST]
<|assistant|>
\end{verbatim}
}

The model output is generated using the following API call:

{\footnotesize
\begin{verbatim}
model.generate(**tokenized_input, max_new_tokens=32)
\end{verbatim}
}

This setup follows the default WildGuard configuration without additional parameters. Since the experiment focuses on prompt-only evaluation, the \texttt{response} field is left empty. If the model outputs \texttt{"harmful request: yes"}, the input prompt is classified as \emph{unsafe}; otherwise, it is considered \emph{safe}.
 
\subsection{Llama-Guard-3 Evaluation}
\label{sec:llama_guard_evaluation}
 
We evaluate the Llama-Guard-3 model using the HuggingFace Transformers framework, following the guidelines provided in its model card. The model output is generated using the following API call:

{\footnotesize
\begin{verbatim}
def moderate(chat):
    input_ids = tokenizer.apply_chat_template(
        chat, return_tensors="pt"
    ).to(device)
    output = model.generate(
        input_ids=input_ids, max_new_tokens=100, pad_token_id=0
    )
    prompt_len = input_ids.shape[-1]
    return tokenizer.decode(
        output[0][prompt_len:], skip_special_tokens=True
    )
\end{verbatim}
}

Since the experiment focuses on prompt-only evaluation, the chat input contains only the prompt:

{\footnotesize
\begin{verbatim}
[
    {"role": "user", "content": "{prompt}"},
]
\end{verbatim}
}

If the model outputs contain \texttt{"unsafe"}, the input prompt is classified as \emph{unsafe}; otherwise, it is considered \emph{safe}.

\subsection{ADRAG Online Latency Benchmark}
\label{sec:adrag_latency_benchmark}
 
We evaluated ADRAG’s end-to-end latency across its two primary components: Guard model inference and Knowledge Base retrieval. Model deployment was performed using a Triton LLM backend with a TensorRT-LLM (v0.16.0) engine. During deployment, the models were optimized with tensor parallelism (degree~1), KV caching, and dynamic batching. All experiments were conducted on a single NVIDIA H100 GPU.  

For retrieval, ADRAG employs the Stereo dual-encoder family (v0.9.5) as the embedding model, which generates 512-dimensional representations for both queries and documents. These encoders constitute the backbone of the retrieval stage, enabling efficient large-scale vector search. The system leverages Faiss-based indexes tuned to domain characteristics, including HNSW graphs (e.g., HNSW64, HNSW128) for low-latency recall, IVF+HNSW hybrids (e.g., IVF256\_HNSW32, IVF65536\_HNSW32) for large corpora, and quantized variants (e.g., OPQ128\_256, IVF2048\_HNSW32, PQ128) for memory efficiency. Runtime parameters such as \texttt{nprobe} (number of clusters searched), \texttt{efSearch} (search depth), and \texttt{efConstruction} (graph connectivity) were tuned to balance recall and latency, with inner product employed as the default similarity metric. Benchmark domains ranged from small (4–16 shards, $\sim$1–10M vectors) to medium (64 shards, 12–16 GiB, $\sim$10–100M vectors).  

\begin{figure}[H]
    \centering
    \includegraphics[width=0.55\linewidth]{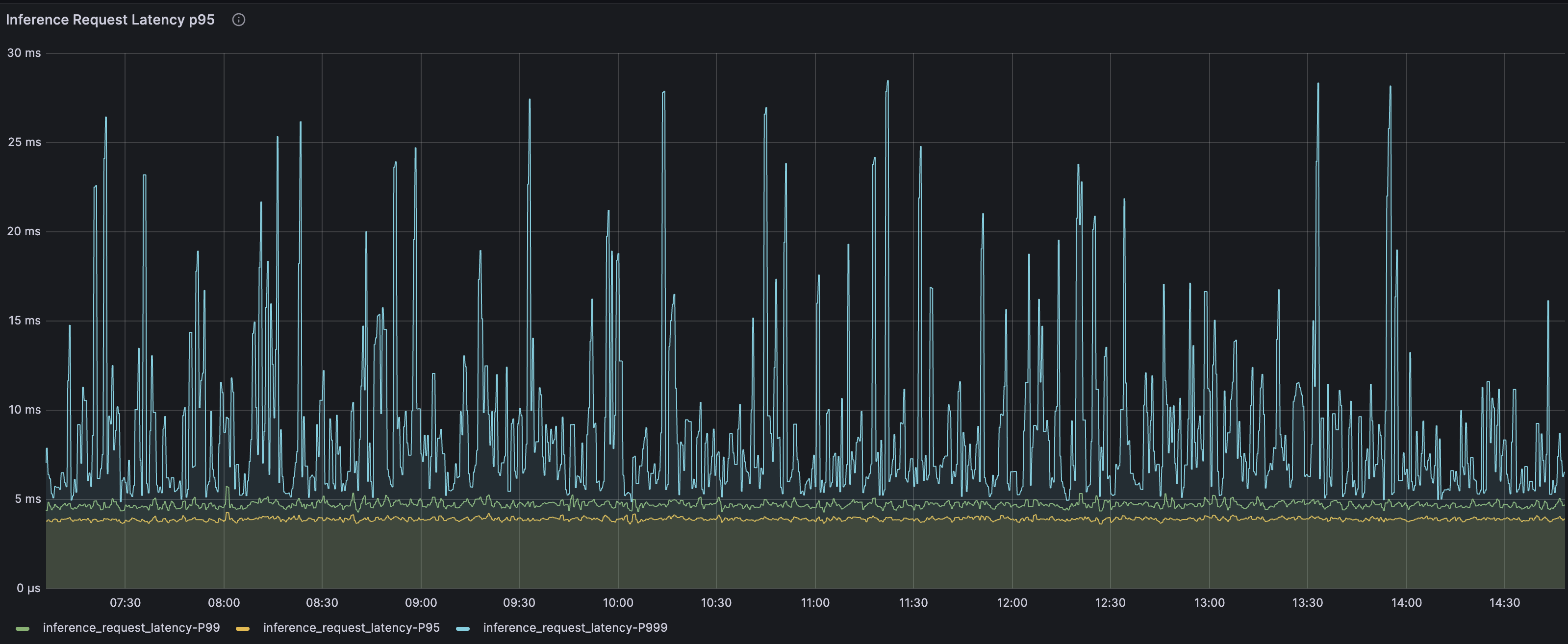}
    \caption{395M RAG model latency (P95 = 4.41\,ms, P99 = 4.43\,ms).}
    \label{fig:latency_adrag_395m}
\end{figure}

\begin{figure}[H]
    \centering
    \includegraphics[width=0.8\linewidth]{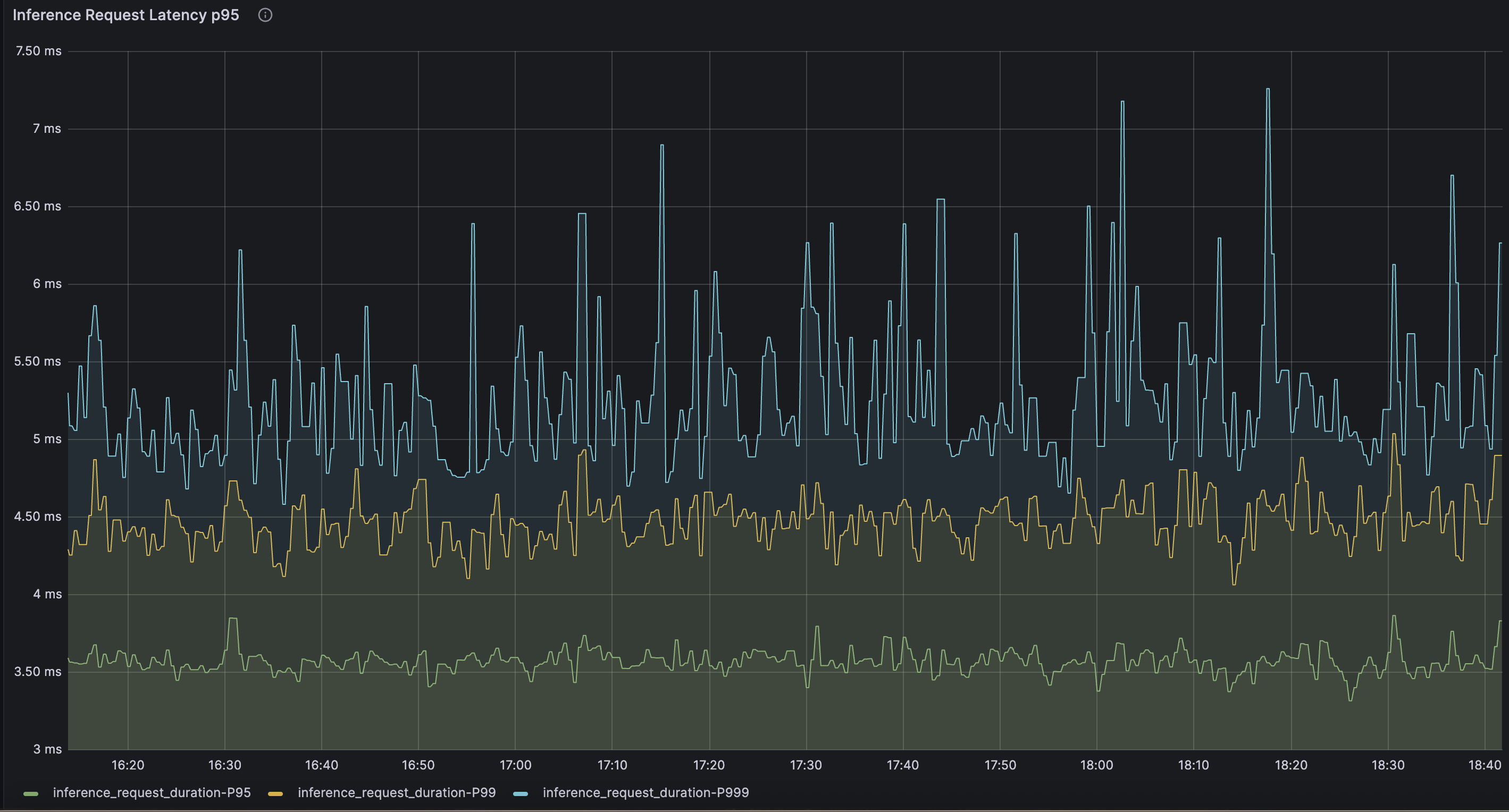}
    \caption{149M RAG model latency (P95 = 2.95\,ms, P99 = 3.07\,ms).}
    \label{fig:latency_adrag_149m}
\end{figure}

\begin{figure}[H]
    \centering
    \includegraphics[width=0.8\linewidth]{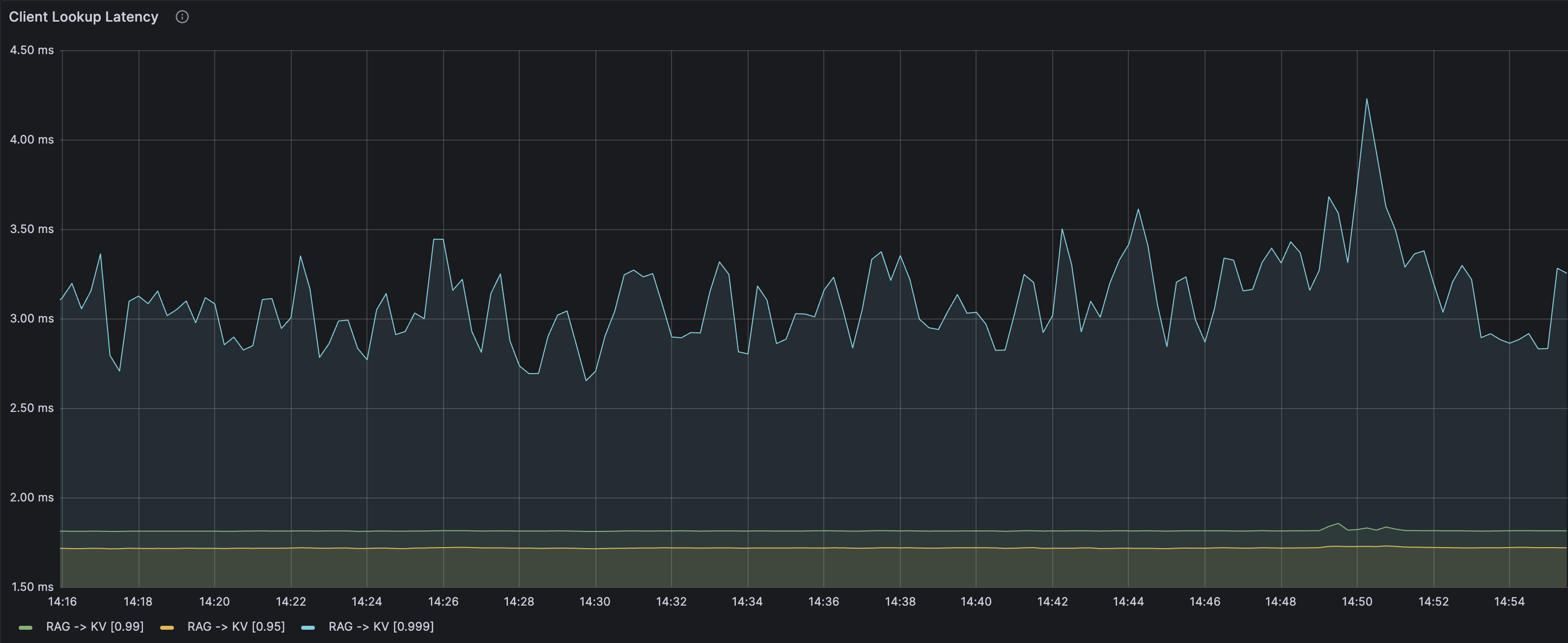}
    \caption{Knowledge Base Lookup latency (P95 = 1.74\,ms, P99 = 1.85\,ms).}
    \label{fig:latency_adrag_rag}
\end{figure}

Under this configuration, ADRAG achieved a throughput of 300 queries per second (QPS). Figures~\ref{fig:latency_adrag_395m}, \ref{fig:latency_adrag_149m}, and \ref{fig:latency_adrag_rag} illustrate the latency distributions obtained from the Grafana monitoring dashboard. The summarized latency results are discussed in Section~\ref{sec:rq1}.

\applefootnote{ \textcolor{textgray}{\sffamily Apple and the Apple logo are trademarks of Apple Inc., registered in the U.S. and other countries and regions.}}

\end{document}